\documentclass[aps,prd,amsmath,amssymb,twocolumn,preprintnumbers,nofootinbib]{revtex4-2}

\DeclareMathAlphabet{\MATHIT}{OT1}{ptm}{m}{it}
\DeclareSymbolFont{Letters}{OML}{ztmcm}{m}{it}
\DeclareSymbolFontAlphabet{\mathNormal}{Letters}
\usepackage[english]{babel}
\usepackage{amsmath}
\usepackage{amsfonts}
\usepackage{amssymb}
\usepackage{amsthm}
\usepackage{enumitem}
\usepackage{dcolumn}
\usepackage{bm}
\usepackage{bbm}
\usepackage{cancel}
\usepackage{tikz}
\usetikzlibrary{calc,shapes,arrows,patterns,decorations.pathmorphing,positioning,automata}
\usepackage{tabularx}
\usepackage{booktabs}
\usepackage{tensor}
\usepackage{scalerel}
\usepackage{xcolor}
\usepackage{mathtools}
\usepackage{float}
\usepackage{xfrac}
\usepackage[pdfborder={0 0 0}]{hyperref}
\definecolor{darkblue}{rgb}{0,0,.5}
\definecolor{darkgreen}{rgb}{0,0.5,.5}
\definecolor{darkyellow}{rgb}{0.5,0.5,0}
\definecolor{fhl}{rgb}{1,0,0}
\hypersetup{colorlinks=true, breaklinks=true, linkcolor=darkblue, menucolor=darkblue, urlcolor=darkblue, linktocpage=true}
\usepackage{geometry}
\usepackage{textpos}
\usepackage{placeins}
\usepackage{relsize}
\usepackage{balance}
\usepackage[titletoc,title]{appendix}
\allowdisplaybreaks
\usepackage{slashed}
\usepackage{natbib}
\usepackage[export]{adjustbox}
\usepackage[most]{tcolorbox}
\tcbuselibrary{skins}

\newcommand{\e}{\mathrm{e}}
\newcommand{\dd}{\mathrm{d}}

\newcommand{\beq}{\begin{eqnarray}}
\newcommand{\eeq}{\end{eqnarray}}

\newcommand{\bmp}{\noindent\begin{minipage}{16cm}}
\newcommand{\emp}{\end{minipage}\vskip 7mm} 

\DeclareMathOperator{\tr}{Tr}

\def\lsim{\mathrel{\rlap{\lower4pt\hbox{\hskip1pt$\sim$}}
    \raise1pt\hbox{$<$}}}                
\def\gsim{\mathrel{\rlap{\lower4pt\hbox{\hskip1pt$\sim$}}
    \raise1pt\hbox{$>$}}}                

\let\originalleft\left
\let\originalright\right
\renewcommand{\left}{\mathopen{}\mathclose\bgroup\originalleft}
\renewcommand{\right}{\aftergroup\egroup\originalright}
\newcommand{\SU}[1]{\operatorname{SU}\left(#1\right)}
\newcommand{\of}[1]{\left(#1\right)}

\newcommand{\sbof}[1]{\Bigl(\Big.#1\Big.\Bigr)}
\newcommand{\sof}[1]{\bigl(\big.#1\big.\bigr)}
\newcommand{\ssof}[1]{(#1)}
\newcommand{\fof}[1]{\left[#1\right]}

\newcommand{\cof}[1]{\left\{#1\right\}}

\newcommand{\sscof}[1]{\{#1\}}
\newcommand{\abs}[1]{\lvert#1\rvert}

\newcommand{\ssabs}[1]{|#1|}

\newcommand{\avof}[1]{\left\langle #1\right\rangle}

\newcommand{\savof}[1]{\big\langle #1\big\rangle}
\newcommand{\ssavof}[1]{\small\langle #1\small\rangle}

\newcommand{\order}{\mathcal{O}}

\definecolor{newgreen}{RGB}{10,100,20}

\iffalse
\definecolor{refcorrcol}{RGB}{200,0,0}
\newcounter{inrefcorr}
\newcommand{\refcorr}[2][]{\stepcounter{inrefcorr}\ifmmode{{\color{refcorrcol}#2}}\else{{\color{refcorrcol}#2}}\fi\addtocounter{inrefcorr}{-1}}
\everymath{\ifnum\value{inrefcorr}>0\color{refcorrcol}\fi}
\else
\newcommand{\refcorr}[2][]{#2}
\fi

\iffalse
\definecolor{refcorrbcol}{RGB}{200,0,0}
\newcounter{inrefcorrb}
\newcommand{\refcorrb}[2][]{\stepcounter{inrefcorrb}\ifmmode{{\color{refcorrbcol}#2}}\else{{\color{refcorrbcol}#2}}\fi\addtocounter{inrefcorrb}{-1}}
\everymath{\ifnum\value{inrefcorrb}>0\color{refcorrbcol}\fi}
\else
\newcommand{\refcorrb}[2][]{#2}
\fi

\definecolor{commentcol}{RGB}{255,0,0}
\newcounter{incomment}
\newcommand{\ucomment}[2][]{\stepcounter{incomment}\ifmmode{{\color{commentcol}#2}}\else{{\color{commentcol}#2}}\fi\addtocounter{incomment}{-1}}

\graphicspath{
{./}{./img/}
}

\newcommand{\re}{\mathrm{Re\,}}

\begin{document}

\title{Confined-deconfined interface tension and latent heat in $\SU{N}$ gauge theory}

\author{Tobias Rindlisbacher}
\email{tobias.rindlisbacher@helsinki.fi}
\affiliation{Department of Physics \& Helsinki Institute of Physics, University of Helsinki, P.O. Box 64, FI-00014 University of Helsinki, Finland}
\author{Kari Rummukainen}
\email{kari.rummukainen@helsinki.fi}
\affiliation{Department of Physics \& Helsinki Institute of Physics, University of Helsinki, P.O. Box 64, FI-00014 University of Helsinki, Finland}
\author{Ahmed Salami}
\email{ahmed.salami@helsinki.fi}
\affiliation{Department of Physics \& Helsinki Institute of Physics, University of Helsinki, P.O. Box 64, FI-00014 University of Helsinki, Finland}

\begin{abstract}
We present high-precision lattice results for the confined-deconfined interface tension and the latent heat of pure SU($N$) gauge theories up to $N=10$ and investigate their asymptotic $N$-dependency. For both quantities we observe the leading $N^2$ behaviour and subleading corrections, with the result for the interface tension $\sigma/T_c^3 = 0.0182(7) N^2 - 0.194(15)$ and for the latent heat $L/T_c^4 = 0.360(6) N^2 - 1.88(17)$.  We use the \emph{mixed phase ensemble} method - where the system is constrained so that half of the volume is in the confined phase and the other half in the deconfined phase - and the interface tension is obtained by measuring the capillary wave fluctuation spectra of the interfaces between the two phases.
The method bypasses supercritical slowing down from which other methods for determining the interface tension suffer, and as a by-product produces accurate estimates of the critical inverse gauge coupling as a function of the inverse temperature. We use the latter to determine the lattice beta function values, required to compute the latent heat from the discontinuity in the average plaquette action across the confined-deconfined transition.   
\end{abstract}

\preprint{HIP-2025-21/TH}

\maketitle

\section{Introduction}\label{sec:intro}

Extensive work using lattice simulation methods has unambiguously shown that non-Abelian gauge field theories have a phase transition from low temperature and density confined phase to high temperature or density deconfined phase, provided the number of fermionic degrees of freedom is not too large (for an overview, see ref.\,\cite{Aarts:2023vsf}).  Pure SU($N$) gauge theories at large $N$ are a particularly interesting case for studying thermodynamics in strongly coupled theories. The transition is of second order at $N=2$ and first order at $N\ge 3$, becoming increasingly strong as $N$ grows. The absence of fermions means that it is possible to obtain precise results with moderate computational effort. At large $N$ this enables studying 
the approach to the 't Hooft limit \cite{tHooft:1973alw} and testing duality relations \cite{Witten:1998zw,Ares:2021ntv}.

The latent heat and the interface tension are the most important quantities characterizing a first order phase transition, since they determine to a large extent the degree of supercooling (or -heating) and the energy release in the phase transition. In the context of cosmology, phase transitions in dark sector SU($N$) extensions of the Standard Model have been analyzed in light of gravitational wave production \cite{Huang:2020crf,Sanchez-Garitaonandia:2023zqz,Ares:2020lbt,Bigazzi:2020avc,Morgante:2022zvc,Zhu:2021vkj}.

In this article we report high-precision results for the latent heat and the interface tension of the deconfinement phase transition in SU($N$) gauge theories at $N=4,5,8$ and $10$.  We unambiguously observe the leading $O(N^2)$ behavior and $O(1)$ subleading terms, as expected in the 't Hooft limit.  At the same time, we obtain the critical lattice couplings as functions of the imaginary time extent ($N_t$) of the lattice, and determine the lattice $\beta$-functions in the relevant coupling constant range.

While the large-$N$ equation of state \cite{Panero:2009tv}, critical temperature $T_c$ 
\cite{Lucini:2012wq}, and the latent heat \cite{Lucini:2005vg} have been studied extensively, the interface tension remains poorly determined \cite{Lucini:2005vg}. The results in this work represent a significant increase in precision compared with the existing literature.
Preliminary results were published in ref.~\cite{Salami:2025iqq}.

We use the {\em mixed phase ensemble} method, where the order parameter, in our case the Polyakov line, is constrained so that half of the lattice is in the confined phase and the other half is in the deconfined phase. With periodic boundary conditions there will be two phase interfaces spanning the system. This method enables us to obtain a precise value of the critical coupling, the interface tension and the latent heat. 

With the mixed phase ensemble the interface tension can be measured using {\em capillary wave theory} (CWT) \cite{Buff:1965zz,RowlinsonWidom}, which states that the magnitude of long-wavelength transverse fluctuations of the interface are inversely proportional to the interface tension. 
It has been applied in fluid dynamics to measure the interface tension both in experiments \cite{RowlinsonWidom,Penfold} and in atomistic simulations \cite{Chacon,Senapati}.
To our knowledge the interface fluctuations were first used to measure the interface tension in a lattice field theory context by Moore and Turok almost 30 years ago \cite{Moore:1996bn}, but it appears to have been used since much less than order parameter distribution (histogram) methods \cite{Binder:1981sa}. The advantage of the mixed phase method is that it does not suffer from supercritical slowing down, as described in Section \ref{sec:mixedphaseruns}, enabling the use of large volume lattices.  We note that essentially similar fluctuation physics underlies the L\"uscher term in QCD Wilson loops \cite{Luscher:1980ac}.

The structure of this paper is the following: in Sec.~\ref{sec:mixedphaseruns} we describe the implementation of mixed phase simulations and the measurement of the critical couplings.  In Sec.~\ref{sec:interfacetension} we discuss the capillary wave method 
for the interface tension measurement, and in Sec.~\ref{sec:latenheat} the lattice $\beta$-function and the latent heat. We conclude in Sec.~\ref{sec:conclusions}.

\section{Mixed phase lattice runs}\label{sec:mixedphaseruns}

\subsection{Lattice setup}\label{ssec:latticesetup}

We consider SU($N$) gauge field theory on a 4-dimensional Euclidean lattice with periodic boundary conditions and $N_x \times N_y \times N_z \times N_t$ lattice sites, where $N_t$ is the number of lattice sites to imaginary time direction.
We use the standard Wilson single-plaquette action~\cite{Wilson:1974sk},
\begin{equation}
    S = -\frac{\beta}{N}\sum_{x; \mu<\nu}\mathrm{Re}\tr\sof{U_{\mu\nu}\of{x}}
    \label{eq:action}\ ,
\end{equation}
with the inverse lattice gauge coupling $\beta = 2N/g^2$ and the plaquette variables being defined by
\begin{equation}
U_{\mu\nu}\of{x}=U_{\mu}\of{x}U_{\nu}\of{x+\hat\mu}U^\dagger_{\mu}\of{x+\smash{\hat\nu}} U^\dagger_{\nu}\of{x}\ .
\end{equation}
The temperature $T$ and the lattice spacing $a$ are connected through the relation
$T = 1/(N_t\,a)$.
As usual, with fixed $N_t$ the temperature is controlled by varying the inverse lattice coupling $\beta$, through the inverse lattice beta function $\mathrm d \log a/\mathrm d \beta$.  The order parameter for the transition 
is the volume average of the Polyakov line,
\begin{equation}
    P_L=\frac{1}{V_s}\sum_{\bar x} \underbrace{\tr\of{ U_{t}(\bar x,0) U_{t}(\bar x,1) \ldots U_{t}(\bar x,N_t -1)}}_{P_L\of{\bar x}}\,,\label{eq:polyakovloop}
\end{equation}
where $\bar x$ is a spatial coordinate vector, $V_s =N_x N_y N_z$ the number of spatial lattice sites, and $P_L\of{\bar x}$ the local Polyakov line observable. In the confined (low-temperature) phase the Polyakov line expectation value vanishes, whereas there are $N$ degenerate deconfined (high-temperature) phases:
\begin{equation}
    \langle P_L \rangle = \left\{ \begin{array}{ll}
        0,   & \mbox{confined} \\
        c\,e^{i 2\pi n/N}, ~~ &
        \mbox{deconfined}
    \end{array} \right.
    \label{eq:phases}
\end{equation}
where $n = 0, \ldots, (N -1)$ and $0<c \in \mathbb{R}$ is a function of $N$, $N_t$, and $\beta$.  

We restrict the imaginary time $N_t \ge 5$ in order to avoid the unphysical bulk phase \cite{Lucini:2005vg}.  Indeed, it turns out that the lattice effects are still too large at $N_t = 5$ for it to be useful for continuum limit extrapolation.

The update algorithm is a combination of a heat bath~\cite{Cabibbo:1982zn,Kennedy:1985nu} sweep followed by four overrelaxation~\cite{Brown:1987rra,deForcrand:2005xr} update sweeps. We use SU(2) subgroup updates for both heat bath and overrelaxation.\footnote{We did not observe significant benefit from the full SU($N$) overrelaxation by de Forcrand and Jahn \cite{deForcrand:2005xr}, presumably due to its algorithmic complexity on GPUs.}
The simulation program uses the \texttt{hila} lattice field theory programming framework \cite{HILA}, allowing the same program to be used on different computing platforms, including GPUs.


\subsection{Order parameter probability distribution}\label{ssec:orderparameterprobabdist}

\begin{figure}[h]
    \centering
  \begin{tikzpicture}[scale=0.7,nodes={inner sep=0}]
    \node (SCHEMFL) at (0,0) {\includegraphics[width=0.7\columnwidth]{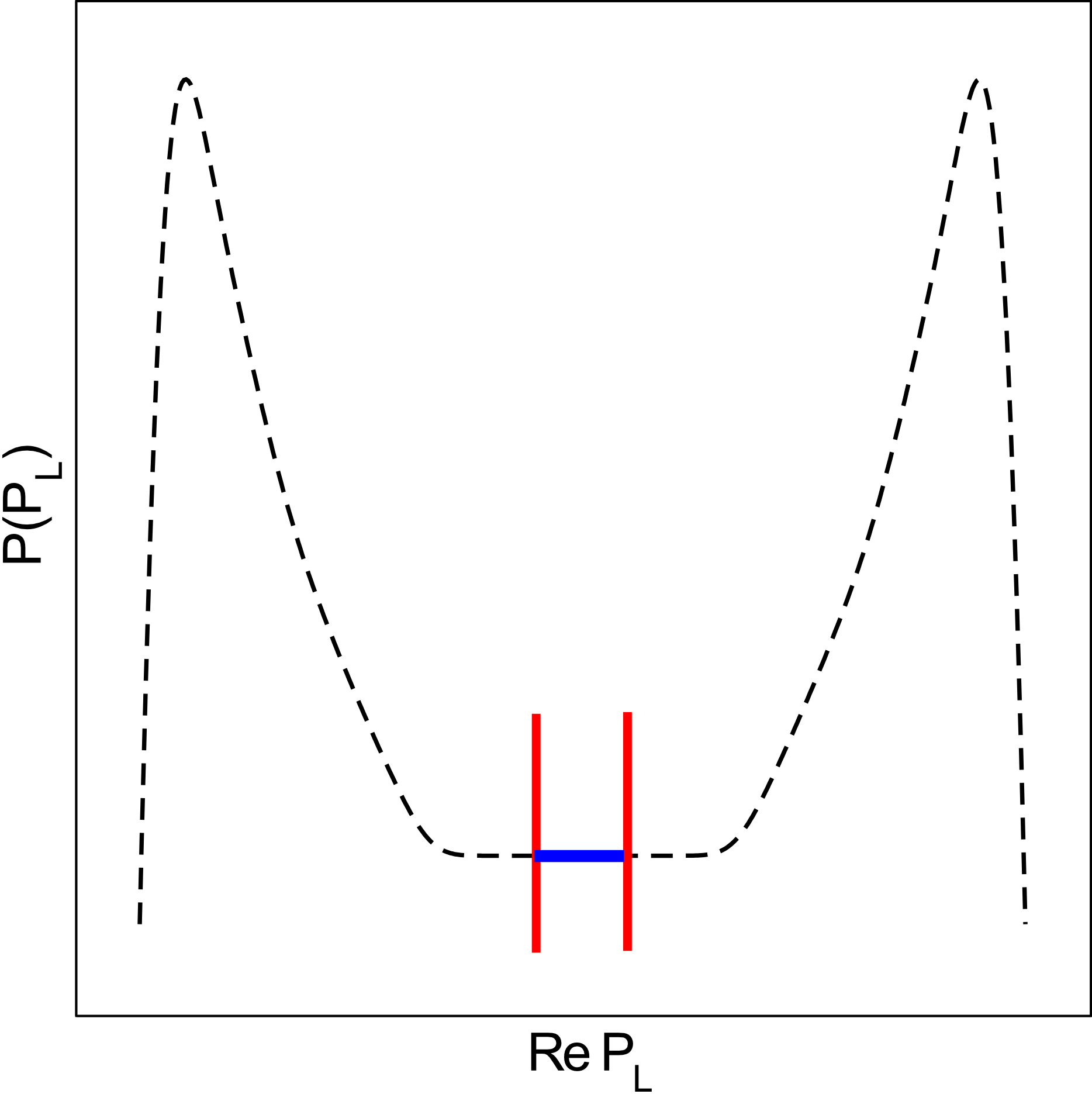}};
    \draw [thick,dotted,gray] (-90pt,100pt) -- (123pt,100pt) node[anchor=west, pos=1.025,scale=1.0] {$P_{\text{max}}$};
    \draw [thick,dotted,gray] (-37.5pt,-66.5pt) -- (123pt,-66.5pt) node[anchor=west, pos=1.025,scale=1.0] {$P_{\text{min}}$};
  \end{tikzpicture}
\caption{Schematic real part of the Polyakov line $\re P_L$ distribution at $\beta_c$ (dashed curve), with restriction to the central region (red lines).}
\label{fig:schematic}
\end{figure}

Let us consider a first-order phase transition with two distinct phases. By definition, at the critical point the two phases have equal probabilities in the large-volume limit.  A schematic order parameter distribution is shown in Fig.~\ref{fig:schematic}: the two high probability peaks in the distribution correspond to the order parameter values characteristic for the two distinct bulk phases, which are bridged by a mixed phase region with suppressed probability.  

\begin{figure}[h]
    \centering
\includegraphics[width=0.75\columnwidth]{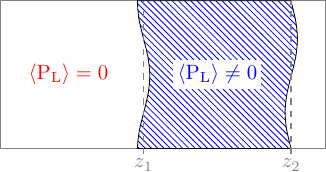}
\caption{Two-phase configuration with two planar interfaces corresponding to the central region of $\re P_L$ distribution in Fig.~\ref{fig:schematic}.}
\label{fig:twophase}
\end{figure}

In Fig.~\ref{fig:twophase} we show a configuration on a periodic volume near the half-way point between the peaks. The two bulk phase domains each occupy half of the volume and are separated by phase interfaces.  Due to the interface tension, the area of the interfaces is minimized.  With the geometry shown in Fig.~\ref{fig:twophase}, the two interfaces settle in planes perpendicular to the $\hat z$-direction. In a sufficiently large volume the central part of the probability distribution becomes flat and the volume fractions occupied by the two phases can therefore change within some range without changing the total free energy. Within these bounds, the two phase interfaces can therefore be considered independent, since they can move freely with respect to each other.

In SU($N$) gauge theory there is one confined phase and $N$ degenerate deconfined phases, all of which have equal probability at the critical point in a sufficiently large periodic volume.  Thus, the system is $N$ times more likely to be deconfined than confined at the critical point. However, in what follows, we restrict the state of the system so that only the confined and the $n=0$ deconfined phase (see Eq.\eqref{eq:phases}) appear.  In this case, the expectation value of the Polyakov line is real, and in the following we use the real part of the Polyakov line, $\re P_L$, as our order parameter. Now, the confined phase with $\langle \re P_L \rangle = 0$ and the deconfined phase $\langle \re P_L \rangle > 0$ have equal probability at the critical point.
 
We note that the slightly skewed shape of the probability peaks in Fig.~\ref{fig:schematic} is caused by bubble-like mixed phase configurations; see ref.~\cite{Moore:2001vf} for a more detailed discussion and \cite{Rummukainen:2025pjj} for an application in SU(8) gauge theory.  These configurations are important for measuring the critical bubble nucleation rate.

The suppressed probability of the mixed states is due to the interface tension $\sigma$: 
\begin{equation}
    \frac{P_{\textrm{min}}}{P_{\textrm{max}}} \approx \exp\left(-\frac{A\,\sigma}{T_c} + \text{finite-\textit{V} corr.}\right)\ ,\label{eq:mixedphasesupprprobab}
\end{equation}
where $A$ is the total area of the interfaces and $P_{\textrm{max}}$ and $P_{\textrm{min}}$ are the maximum and minimum values of the probability distribution of the order parameter, see Fig.~\ref{fig:schematic}. This property is often used to measure the interface tension in lattice simulations. If the interface tension is large, the probability of the mixed phase may be very strongly suppressed, possibly by tens of orders of magnitude. In these cases, standard Monte Carlo sampling will not visit the mixed states. The probability barrier can be overcome with various modified sampling methods: multicanonical~\cite{Berg:1992qua}, Wang-Landau~\cite{Wang:2000fzi}, density of states~\cite{Bennett:2024bhy}, or applications of Jarzynski's theorem~\cite{Caselle:2016wsw}. Multicanonical methods have been used very successfully to study phase transitions in electroweak-like theories on the lattice, see for example~\cite{Kajantie:1995kf}.  However, while modified sampling methods help to overcome the dominant part of the probability suppression, 
these methods still suffer from a milder form of supercritical slowing down as the volume becomes large: there is a condensation barrier for the initial formation of the mixed phase bubbles~\cite{nussbaumer08,Hallfors:2025key}. 

We note here that the interface tension can also be measured from the components of the energy-momentum tensor, as was done in early work in SU(3) gauge theory \cite{Kajantie:1989xk}. However, the energy-momentum tensor on the lattice is an ultraviolet divergent quantity and its non-perturbative renormalization for the interface tension has not been attempted.

\subsection{Constrained mixed phase \refcorr[]{method} and the critical coupling}\label{ssec:constrainedmixedphase}

\begin{figure*}[tb]
    \begin{minipage}[t]{0.32\linewidth}
    \vspace{0pt}
    \centering
    \begin{tikzpicture}[scale=1,nodes={inner sep=0}]
       \node[anchor=south east] at (-0.5pt,200pt) {\includegraphics[width=154pt]{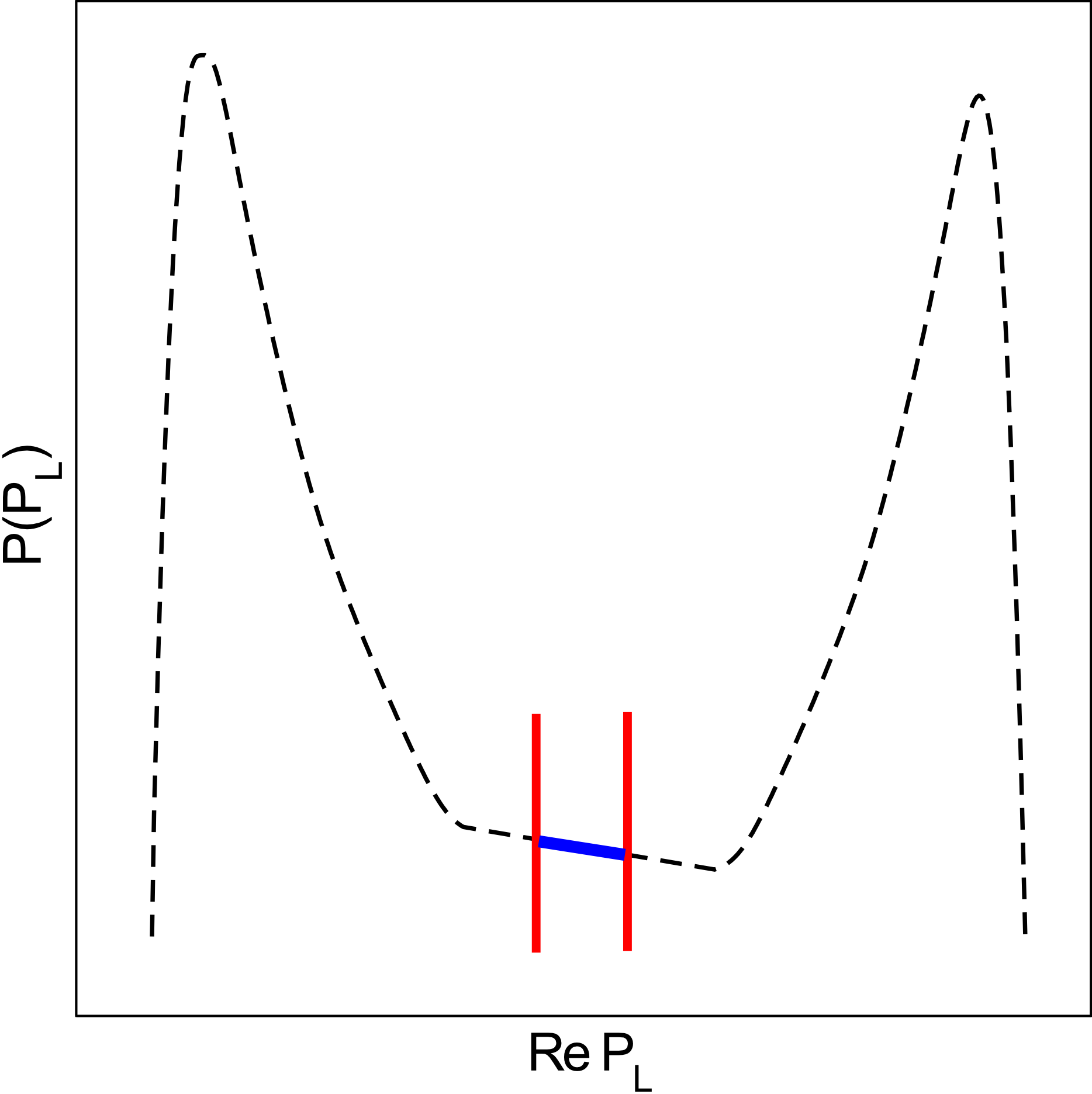}};
       \node[anchor=south east] at (-57pt,320pt) {$\beta<\beta_c$};
       \node[anchor=south east] at (0,30pt) {\includegraphics[height=160pt,keepaspectratio,right]{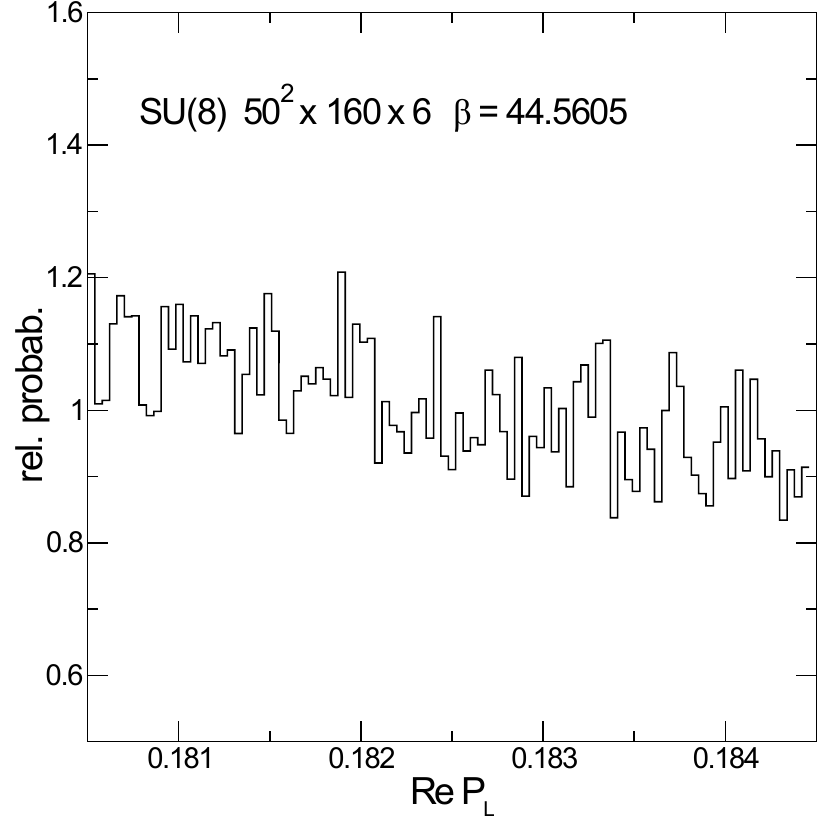}};
    \end{tikzpicture}
    \end{minipage}\hfill
    \begin{minipage}[t]{0.32\linewidth}
    \vspace{0pt}
    \centering
    \begin{tikzpicture}[scale=1,nodes={inner sep=0}]
       \node[anchor=south east] at (-0.5pt,200pt) {\includegraphics[width=154pt]{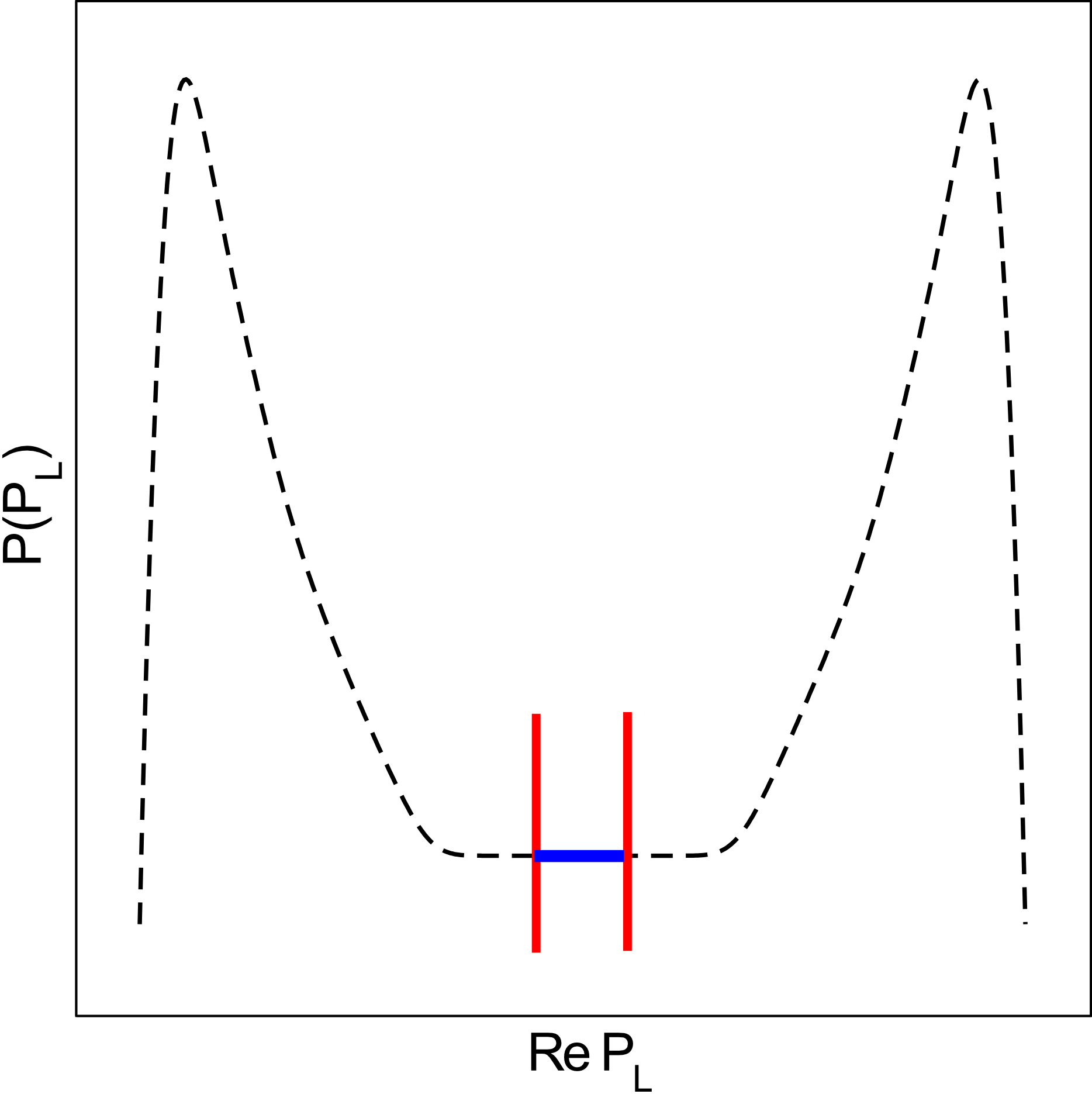}};
       \node[anchor=south east] at (-57pt,320pt) {$\beta=\beta_c$};
       \node[anchor=south east] at (0,30pt) {\includegraphics[height=160pt,keepaspectratio,right]{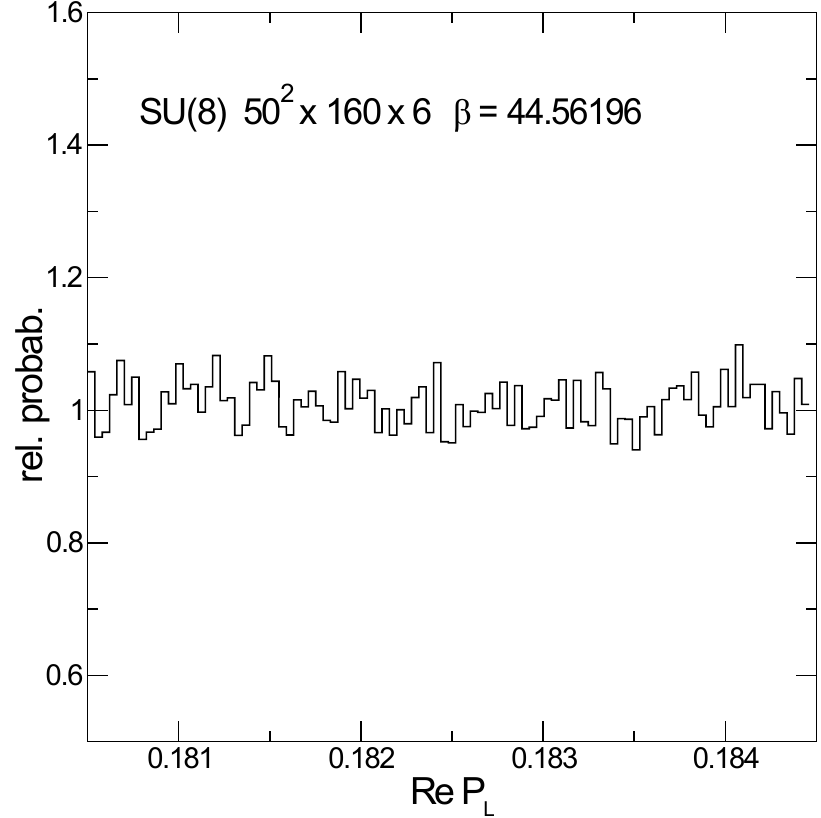}};
    \end{tikzpicture}
    \end{minipage}\hfill
    \begin{minipage}[t]{0.32\linewidth}
    \vspace{0pt}
    \centering
    \begin{tikzpicture}[scale=1,nodes={inner sep=0}]
       \node[anchor=south east] at (-0.5pt,200pt) {\includegraphics[width=154pt]{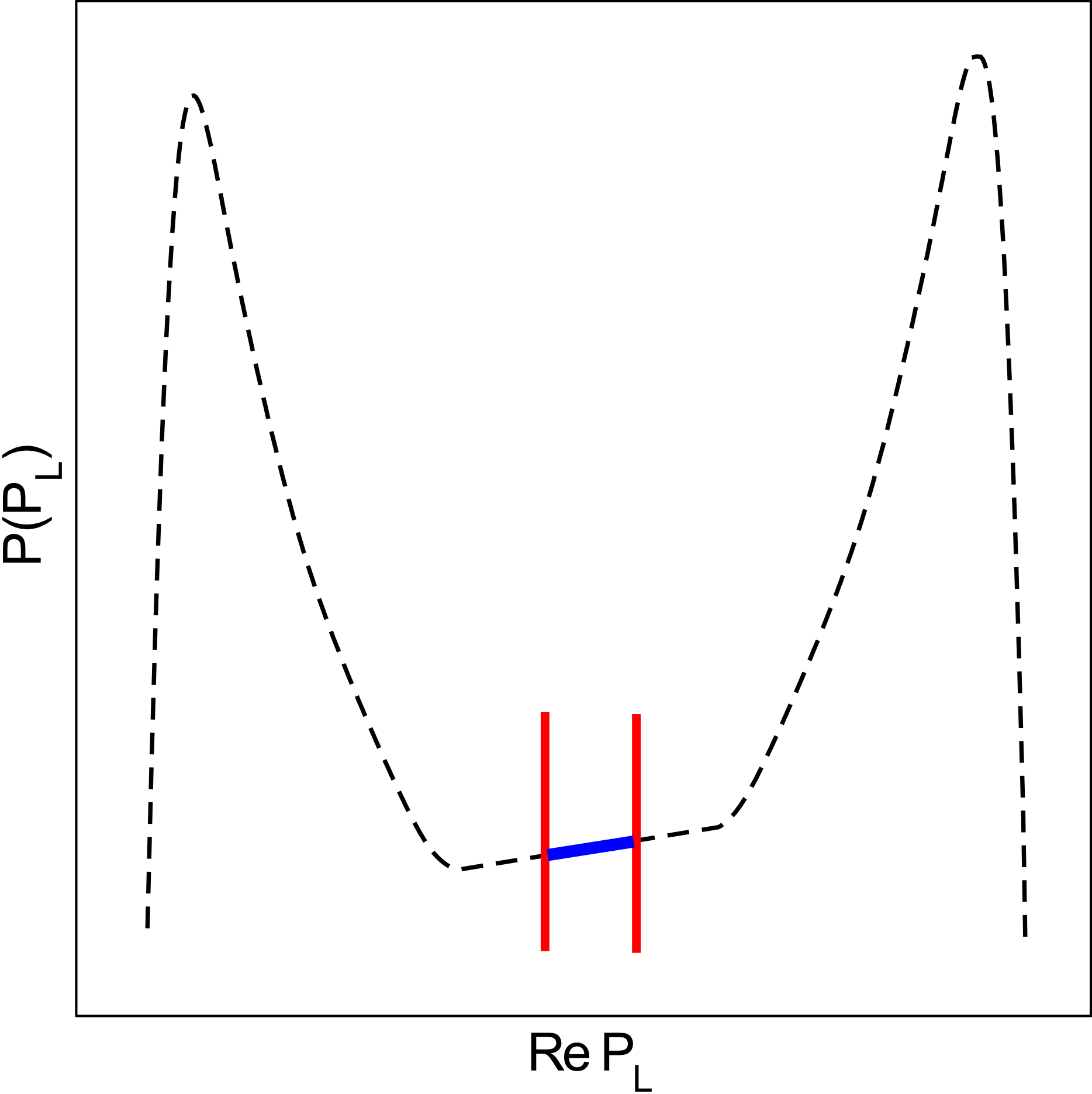}};
       \node[anchor=south east] at (-57pt,320pt) {$\beta>\beta_c$};
       \node[anchor=south east] at (0,30pt) {\includegraphics[height=160pt,keepaspectratio,right]{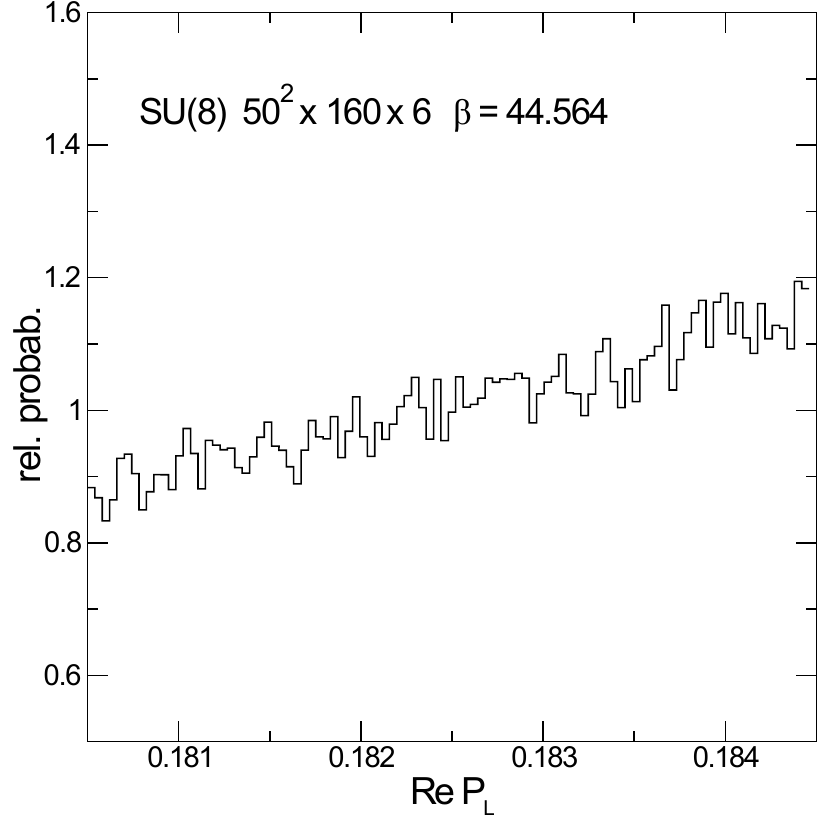}};
    \end{tikzpicture}
    \end{minipage}\\[5pt]
    \caption{\refcorr[Restricted average real part of the Polyakov line $\re P_L$ probability distributions of SU(8) gauge theory on a $6\times 50^2 \times 160$ lattice slightly below, at, and above the critical $\beta$.  In the simulation $\re P_L$ was restricted to range $0.1805 \le \re P_L \le 0.1845$.]{\emph{Top row:} Schematics of how the order parameter probability distribution is expected to look globally and restricted to a small interval at the center of the distribution (marked by the vertical red lines) at values of the inverse gauge coupling $\beta$ slightly below (left), equal to (center), and slightly above (right) its pseudo critical value $\beta_c$. \emph{Bottom row:} Measured probability distributions of the real part of the Polyakov line $\re P_L$  of SU(8) gauge theory on a $6\times 50^2 \times 160$ lattice slightly below, at, and above the critical $\beta$. In the simulation $\re P_L$ was restricted to range $0.1805 \le \re P_L \le 0.1845$.}}
    \label{fig:histsforbetac}
\end{figure*}

\refcorr[The mixed phase method \cite{Moore:1996bn,Moore:2000jw} avoids the supercritical slowing down by sampling only a part of the order parameter distribution, a narrow range around the center of the distribution, as shown in Fig.~\ref{fig:schematic}. This is achieved by initiating the simulation with a suitable mixed-phase configuration and then rejecting updates that would move the value of the average Polyakov line beyond the boundaries of the admissible range. The outcome is that the system remains in the mixed phase, as shown schematically in Fig.~\ref{fig:twophase}. With a sufficiently elongated lattice in the $z$ direction, the interfaces minimize their area by spanning across the ($x,y$) plane and are well separated.]{In this work, we employ the mixed phase method \cite{Moore:1996bn,Moore:2000jw} to accurately determine both the critical coupling $\beta_c$ and the interface tension $\sigma$. In contrast to the methods mentioned in the previous section, which use Eq.~\eqref{eq:mixedphasesupprprobab} to extract the interface tension from measurements of the relative suppression of mixed phase configurations compared to pure phase ones in the thermal ensemble, the mixed phase method aims at extracting the interface tension from measurements of the distribution of surface deformation spectra of the phase interfaces in the ensemble of a specific type of mixed phase configurations, schematically shown in Fig.~\ref{fig:twophase}. The method thus avoids supercritical slowing down associated with the sampling of the full order parameter probability distribution, since only a relatively narrow range around the center of the distribution needs to be sampled, shown in Fig.~\ref{fig:schematic} as blue interval. We achieve this by initiating the simulation with a suitable mixed-phase configuration and then rejecting updates that would move the value of the average Polyakov line beyond the boundaries of the admissible range. This \emph{restricted sampling} with respect to the value of the order parameter satisfies detailed balance and is ergodic on the ensemble of configurations that satisfy the constraint.

We note that it is crucial to have a lattice which is sufficiently elongated in $z$ direction, so that the phase interfaces minimize their area by spanning across the ($x,y$) plane and are always well separated in the $z$ direction to be considered noninteracting. The overall system size should be large enough so that the value of $P_L$ defined in Eq.~\eqref{eq:polyakovloop} adequately reflects the volume fraction occupied by the deconfined phase, i.e. the volume should be large enough so that bulk fluctuations of the local order parameter average out in Eq.~\eqref{eq:polyakovloop}.}

\refcorr[We prepare the initial configuration]{Assuming that we have a sufficiently good initial estimate of the pseudo-critical value $\beta_c$ of the inverse lattice gauge coupling $\beta$, we prepare the initial mixed phase configuration} using simulations with \refcorr[the inverse lattice gauge coupling $\beta$]{$\beta$} slightly \refcorr[above and below the pseudo-critical value $\beta_c$]{above and below $\beta_c$} \refcorr[on the two sides]{in the two halves} of the lattice given by $z<N_z/2$ and $z\geq N_z/2$, respectively. This automatically gives rise to a two-phase configuration \refcorr[]{as sketched in Fig.~\ref{fig:twophase} (up to a translation s.t. $z_2=N_z$)}.  We also ensure that the deconfined phase Polyakov line expectation value is along the positive real axis; because of the very \refcorr[large]{strong} metastability \refcorr[]{of the $N$ degenerate deconfined vacua in large systems}, the expectation value remains real over any realistic simulation time, justifying the use of the real part of the Polyakov line as the order parameter.  After \refcorr[the initial configuration has been generated]{the system has settled in a mixed-phase configuration}, we set $\beta$ to $\beta_c$ \refcorr[]{for all $z$} and impose the above mentioned restriction on the value of the average Polyakov line, that
\begin{equation}
\re P_{L}\in\fof{(1/2-c_w)\,P_0,(1/2+c_w)\,P_0}\ ,\label{eq:polyakovlineconstraint}
\end{equation}
where $P_0=\abs{\avof{P_L}_{\text{decon.}}}$ and the relative interval width, $c_w$, is set to a small \%-value. If the range is too narrow, the update rejection rate grows too large, and the simulations are inefficient. If the range is too wide, the simulation might encounter metastabilities, because the system starts to experience the tails of the two peaks in the probability distribution for $\re P_L$ sketched in Fig.~\ref{fig:schematic} and will be attracted by the boundaries of the Polyakov loop range.   

\refcorr[At]{As mentioned earlier, at} the critical value of the inverse gauge coupling $\beta$, both bulk phases have equal probability weight, or equivalently, equal free energy density. \refcorr[]{The average probability weight of mixed phase configurations in which the two phases occupy certain volume fractions (corresponding to a certain value of the order parameter $P_L$) does then not directly depend on how much volume each phase occupies, but rather on how large the total interface area between the two phases is.} This implies that in \refcorr[the]{our} mixed phase setup \refcorr[]{(cf. Fig.~\ref{fig:twophase})}, the volume fractions of the two phases can vary (within the constraint given by Eq.~\eqref{eq:polyakovlineconstraint}) without changing the probability, resulting in a flat probability distribution of the restricted order parameter. 

However, even a small deviation of $\beta$ away from $\beta_c$ is sufficient to \refcorr[change the probability of the bulk phases and to tilt the distribution]{render the free energy densities in the two bulk phases different and cause the order parameter distribution to tilt, as sketched in the upper panels of Fig.~\ref{fig:histsforbetac}}.  \refcorr[An example of this is shown in Fig.~\ref{fig:histsforbetac}]{The lower panels of Fig.~\ref{fig:histsforbetac} show corresponding simulation data} for SU(8) on a lattice of size $N_x\times N_y\times N_z\times N_t=50^2\times 160\times 6$.  We use this property to accurately determine $\beta_c$: \refcorr[]{we initialize a mixed-phase simulation as described above, using an initial estimate for $\beta_c$, obtained from the literature~\cite{Lucini:2005vg,Lucini:2012wq} or from standard small-volume simulations. Once the system has settled in a mixed phase configuration and the simulation is in the restricted sampling stage, we monitor the order parameter distribution and slightly increase or decrease $\beta$, depending on whether the distribution indicates that the confined or deconfined phase has a larger free energy density for the current value of $\beta$. The value of $\beta$ for which the order parameter distribution appears flat is then identified with $\beta_c$.}
The final value of $\beta_c$ is obtained by reweighting and error analysis is done using the jackknife method. \refcorr{To be more specific, we split the sequence $M$ of Monte Carlo measurements into $N_{\text{JK}}$ equally sized bins $b_i$, $i=1,\ldots,N_{\text{JK}}$, and define the corresponding jackknife sets $B_i=M\setminus b_i$. We then reweight the order parameter probability distribution for each jackknife set $B_i$ in $\beta$ until it appears flat. As flatness criterion we require that the total probability for $\re P_L<P_0/2$ equals the total probability for $\re P_L>P_0/2$, where $P_0/2$ is the midpoint of the order parameter interval from Eq.~\eqref{eq:polyakovlineconstraint}. This procedure yields a list $\cof{\beta_{c,1},\ldots,\beta_{c,N_{\text{JK}}}}$ of jackknife estimates for the critical $\beta$-value. The final value of $\beta_c$ and its error are then obtained, respectively, as 
\begin{equation}
\beta_c=\frac{1}{N_{\text{JK}}}\sum_{i=1}^{N_{\text{JK}}}\beta_{c,i}
\end{equation}
and
\begin{equation}
\operatorname{Err}\of{\beta_c}=\sqrt{\frac{N_{\text{JK}}-1}{N_{\text{JK}}}}\sum_{i=1}^{N_{\text{JK}}}\,\of{\beta_{c,i}-\beta_{c}}^2\ .
\end{equation}
} 

\begin{figure}[h]
    \centering
    \includegraphics[width=0.9\linewidth]{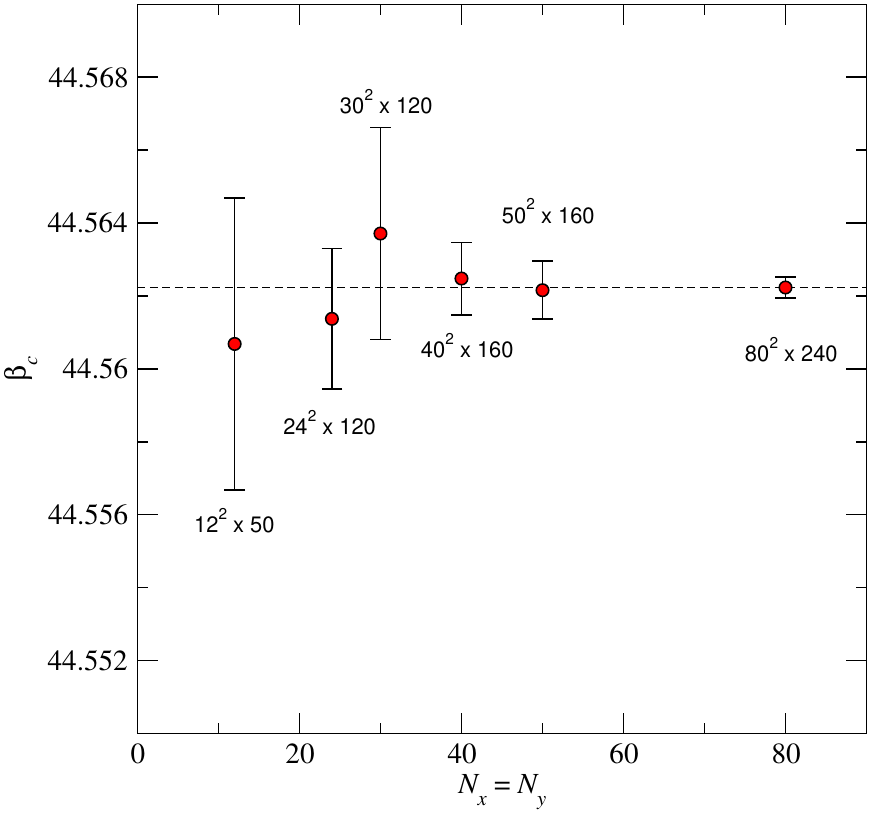}
    \caption{Volume dependence of the critical coupling $\beta_c$ on SU(8), $N_t=6$ lattices.}
    \label{fig:su8finiteV}
\end{figure}

We note that this method of determining $\beta_c$ has exponential finite volume effects, expected to vanish as $\propto \exp(-L/\xi)$, where $\xi$ is some correlation length of the system and $L$ the linear size. This is due to the exponentially suppressed finite volume effects in bulk phase probabilities.
As an example, Fig.~\ref{fig:su8finiteV} shows the results of the determination of $\beta_c$ in SU(8) gauge theory on $N_t=6$ lattices. No systematics in finite volume behavior is observed.
Thus, in each case, we determine $\beta_c$ from the largest volume used for each $N$ and $N_t$. All volumes satisfy the inequalities \refcorrb[$N_t \le 10\,N_{x,y} \ll N_z$]{$10\,N_t \le N_{x,y} \ll N_z$}, with the exception of the $N_t=10$ cases, which have been added to improve the accuracy of the lattice beta function required to determine the latent heat in Sec.~\ref{sec:latenheat}, and the single SU(16), $N_t=6$ lattice. This point was included as a consistency check, and \refcorr[it is]{is} not used in our main analysis. The simulation volumes and $\beta_c$ values are reported in Table~\ref{tab:results}.

\section{Interface tension}\label{sec:interfacetension}

Because the lattices are elongated in the $z$ direction, the two interfaces between the confined and the deconfined phase will always span across the $(x,y)$-plane. Let us consider the idealized case of an infinitely thin interface of size $L^2$ with periodic boundary conditions. Assuming there are no overhangs, the interface is described by the height function $z(x,y)$. The energy of the interface is proportional to the area:
\begin{align}
H &= \sigma \int_0^L \mathrm d x\,\mathrm d y\, \sqrt{1 + |\nabla z|^2}
\label{eq:Hsurf}
\end{align}
where the interface tension $\sigma$ restricts the thermal fluctuations of $z(x,y)$. In long wavelengths the spectrum of $z(x,y)$ fluctuations 
approach the {\em capillary wave} limit \cite{Buff:1965zz, RowlinsonWidom}, where the amplitude of the fluctuations is much smaller than the wavelength. In this limit Eq.~\eqref{eq:Hsurf} can be linearized, and the components of the 2-dimensional Fourier transform of $z(x,y)$ obey
\begin{equation}
    \langle |\hat z(k)|^2 \rangle = \frac{T}{\sigma L^2 k^2} = \frac{T}{4\pi^2 \sigma(n_x^2+n_y^2) } ~,
    \label{eq:zhat}
\end{equation}
where $k = 2\pi n/L$ is a 2-dimensional wave vector along the $(x,y)$-plane, and the expectation value is over the ensemble of configurations.  For the derivation and corrections to Eq.~\eqref{eq:zhat} see Appendix \ref{app:fluctuations}. It turns out that due to the corrections to the capillary wave picture, the limit $k \rightarrow 0$ has to be taken in Eq.~\eqref{eq:zhat}. Indeed, the value of the interface tension itself is $k$-dependent, and only the limit $k\rightarrow 0$ gives us a uniquely defined result. In a realistic field theory the interface is not infinitely thin; however, in the limit of long wavelengths ($k\rightarrow 0$) the capillary wave picture becomes valid.

Thus, in the constrained mixed phase ensemble the interface tension can be obtained by (a) locating the interface surface, $z(x,y)$, for each configuration (for periodic lattices, the two interface surfaces $z_i(x,y)$, $i=1,2$ ), (b) calculating the Fourier transform $\hat z(k)$ and (c) extrapolating the result with Eq.~\eqref{eq:zhat} to $k\rightarrow 0$.  In field theory simulations this method has been used to measure $\sigma$ in the SU(2)-Higgs model \cite{Moore:1996bn,Moore:2000jw} and, with some modifications, in atomistic condensed matter simulations (see e.g. ref.~\cite{Chacon}).

We note here that the capillary wave theory, Eq.~\eqref{eq:zhat}, implies logarithmic broadening of the averaged interface width \cite{RowlinsonWidom}:
\begin{equation}
    \langle w^2 \rangle = \frac{T}{2\pi\sigma} \log\of{L\,\mu}
\end{equation}
where $\mu$ is an ultraviolet cutoff scale, and
\begin{equation}
    w^2 = \int_A (z(x,y) - \bar z)^2 
    \textrm{~with~}
    \bar z = \int_A z(x,y),
\end{equation}
where we have denoted $\int_A = \frac1{L^2}\int \mathrm dx\,\mathrm dy$.  This property has been used to analyze interface tension, both in experiments and in simulations \cite{Buff:1965zz,RowlinsonWidom,Penfold,Senapati}.

Returning to the SU($N$) gauge theory, we define the local position of the interface $z(x,y)$, often called the \emph{Gibbs surface}, from the real part of the Polyakov line field, $\re P_L(x,y,z)$:
for each configuration and for each $(x,y)$ we search for two values (for two interfaces) of $z$ where $\re P_L(x,y,z)$ crosses a threshold value of $P_0/2$, where $P_0=\abs{\avof{P_L}_{\text{decon.}}}$ as in Eq.~\eqref{eq:polyakovlineconstraint}.  The precise value of the threshold is not important, but choosing a value near $P_0/2$ reduces the effect of bulk fluctuations.


\begin{figure*}[htb]
    \centering
    \includegraphics[width=0.9\linewidth]{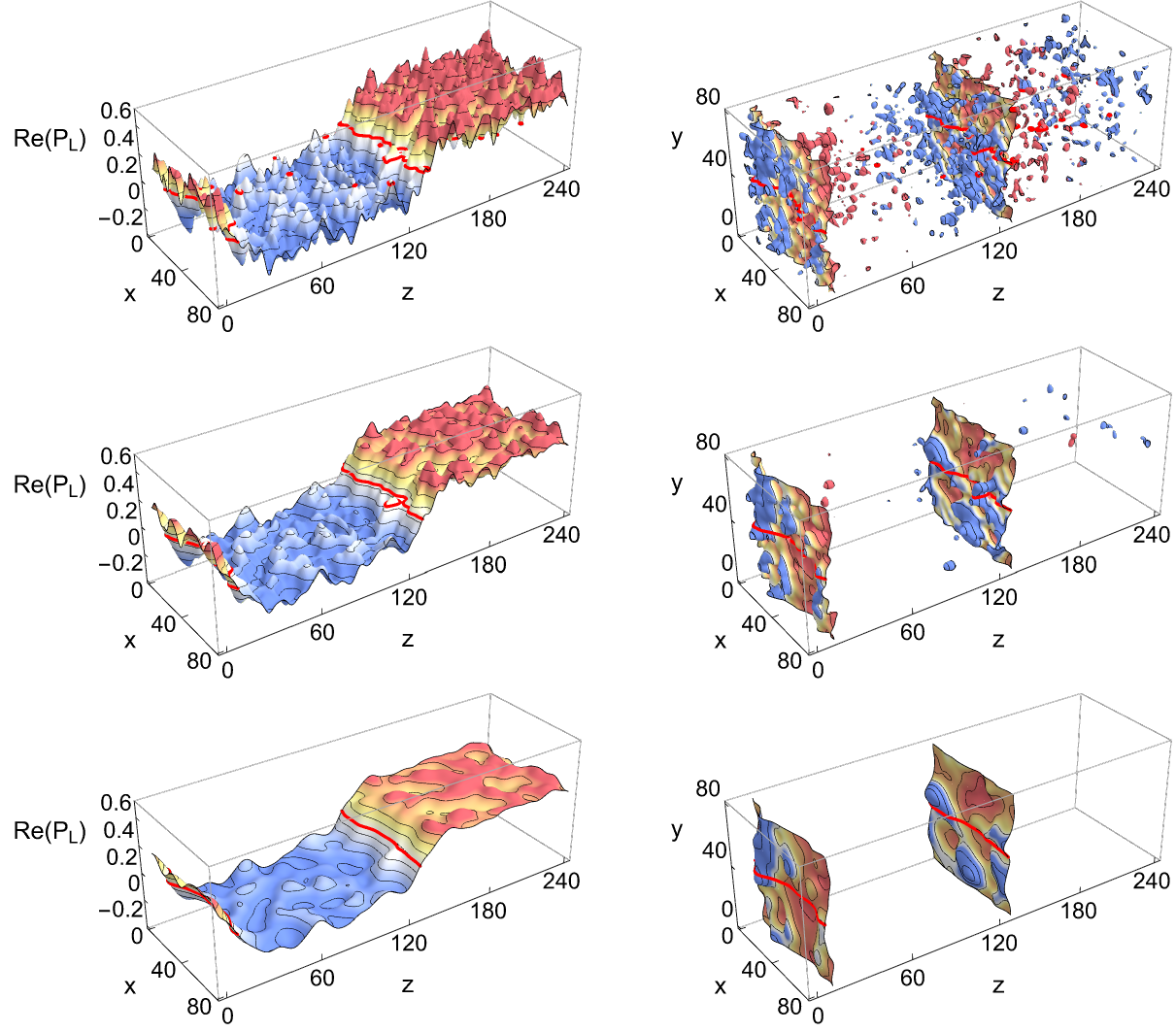}
    \caption{Capturing the confined-deconfined interface from a configuration in coexisting-phase-restricted simulation of SU(8) gauge theory on a $80^2\times 240\times 6$ lattice at $\beta=44.56196$. 
    \emph{Left column:} real trace of Polyakov loop in the $x$-$z$-plane at $y=40$ after $n_s=$10 (top), 20 (middle), and 60 (bottom) repeated convolutions with the smearing kernel from Eq.~\eqref{eq:smearingkernel}. The red lines mark the detected phase boundaries in the $x$-$z$-plane at $y=40$, based on the chosen deconfined threshold value for the Polyakov loop. 
    \emph{Right column:} phase boundary surfaces obtained by performing the deconfined threshold value detection in the Polyakov loop field over the whole spatial volume instead of over a slice of constant $y$ coordinate. The red lines are the same as in the corresponding panels in left column. 
    After 10 smearing steps, UV noise in the local Polyakov loop observable is still significant and prevents a unique detection of the $x$-$y$-spanning phase boundaries. With increasing numbers of smearing steps, the noise gradually disappears and the two $x$-$y$-spanning phase boundaries can be uniquely mapped out and parametrized as functions $z_{i}\of{x,y}$, $i=1,2$.}
    \label{fig:smearing_comp}
\end{figure*}

The bare $P_L(x,y,z)$ is a very noisy observable and its measurement-by-measurement fluctuations can be much larger than the difference between confined and deconfined phase averages. Thus, as such $P_L(x,y,z)$ cannot be used to locate the interface.  A simple way to reduce the noise is to perform recursive smearing, 
\begin{subequations}\label{eq:isotropicsmearing}
\begin{equation}
    P_L^{(n+1)}(\bar x) = \sum_{\bar{y}} S(\bar x,\bar y) P_L^{(n)}(\bar y)\ ,
\end{equation}
with $P_{L}^{(0)}\of{\bar{x}}=P_{L}\of{\bar{x}}$ and smearing kernel
\begin{equation}
    S(\bar x,\bar y) = \frac{1}{1+6\,\rho}\left(\delta_{\bar x,\bar y}+\rho\sum_{i=1}^{3}(\delta_{\bar x+\hat{i},\bar y}+\delta_{\bar x -\hat{i},\bar y})\right)\ .\label{eq:smearingkernel}
\end{equation}
\end{subequations}
Following ref.~\cite{Moore:1996bn}, we implement on top of $n_s$ steps of isotropic smearing with Eq.~\eqref{eq:isotropicsmearing}, also additional smearing in $z$-direction only, by applying the smearing kernel
\begin{equation}
    S_z(\bar x, \bar y) = \frac{1}{1+2\rho} \left[\delta_{\bar x,\bar y} + 
        \rho\,(\delta_{\bar x + \hat e_z,\bar y} + \delta_{\bar x - \hat e_z,\bar y}) \right]
\end{equation}
$n_{s,z}$ times.  

The smearing rapidly removes short-wavelength fluctuations. After $n_s$ isotropic smearing steps the 3-dimensional momentum space Polyakov line is approximately
\begin{equation}
    P_L^{(n_s)}(\bar q) \approx P_L(\bar q)
     \,e^{-\alpha_{\rho} n_s\,a^2 q^2}\ ,\label{eq:plmodesmearingeffect}
\end{equation}
with $\alpha_{\rho}=\rho/\of{1+6\,\rho}$. For a precise analysis of the smearing see Appendix \ref{app:smearingkernel}.
The smearing efficiently reduces the amplitude of fluctuations with wavelength shorter than $\sim a \sqrt{\alpha_{\rho} n_s}$.
Too few smearing steps do not reduce the order parameter fluctuations sufficiently, leading to spurious dislocations in the measurements of the interface location, as illustrated in Fig.~\ref{fig:smearing_comp}.  Increasing the smearing level makes the noise disappear while preserving the long-wavelength structure, unless the number of smearing steps grows excessive. We typically use multiple smearing levels $n_s$ in the range 5 \ldots 160, with $\rho = 0.9$, and the surface location $z(x,y)$ is measured with several smearing levels. The $z$-direction smearing turns out to be less significant, and we keep a fixed value $n_{s,z}=10$.


\setlength{\tabcolsep}{0.3em}
\begin{table}[h]
    \small
    \centering
    \begin{tabular}{c | c | c | c | l | l | l }
        $N$ & $N_t$ & $N_{x,y}$ & $N_z$ & \multicolumn{1}{c |}{$\beta_c$} & \multicolumn{1}{c |}{$\sigma/T_c^3$} & \multicolumn{1}{c }{$L/T_c^4$} \\
    \hline \hline
        \phantom{0}4 & \phantom{0}5 & \phantom{0}80 & 320 & \phantom{0}10.637765(83) & 0.1661(38)  & \phantom{0}5.58(3) \\
          & \phantom{0}6 & \phantom{0}80 & 200 & \phantom{0}10.79191(11)  & 0.1383(29)  & \phantom{0}4.84(2) \\
          & \phantom{0}7 & 100 & 400 & \phantom{0}10.94215(25) & 0.1285(29)  & \phantom{0}4.55(4) \\
          & \phantom{0}8 & 140 & 400 & \phantom{0}11.08443(35) & 0.1213(31)  & \phantom{0}4.44(7) \\
          & 10 & \phantom{0}60 & 480 & \phantom{0}11.33998(57) &             &          \\ 
          & $\infty$& &   &              & 0.0997(76)  & \phantom{0}3.8(2) \\
        \hline
        \phantom{0}5 & \phantom{0}5 & \phantom{0}80 & 240 & \phantom{0}16.87628(15)  & 0.3716(80)  & 10.17(3) \\
          & \phantom{0}6 & 160 & 480 & \phantom{0}17.11085(11) & 0.3192(23)  & \phantom{0}8.95(4) \\
          & \phantom{0}7 & 100 & 400 & \phantom{0}17.34267(21) & 0.2989(76)  & \phantom{0}8.44(5) \\
          & \phantom{0}8 & 120 & 360 & \phantom{0}17.56119(12) & 0.2935(49)  & \phantom{0}8.24(9) \\
          & 10 & 50 & 400 & \phantom{0}17.9577(11)  &             &          \\ 
        & $\infty$& &     &              & 0.258(11)   & \phantom{0}7.2(2) \\

        \hline
        \phantom{0}8 & \phantom{0}5 & \phantom{0}50 & 180 & \phantom{0}43.98229(34)  & 1.434(35) & 28.03(5) \\
          & \phantom{0}6 & \phantom{0}80 & 240 & \phantom{0}44.56196(48)  & 1.182(15) & 24.84(6) \\
          & \phantom{0}7 & \phantom{0}80 & 240 & \phantom{0}45.13525(76)  & 1.136(21) & 24.0(2)\\
          & \phantom{0}8 & 100 & 400 & \phantom{0}45.67833(49) & 1.099(21) & 23.3(3) \\
          & 10 & \phantom{0}30 & 240 & \phantom{0}46.6582(32)  &           &           \\ 
        & $\infty$& &       &            & 0.994(47) & 21.5(4) \\
        \hline
        10 & \phantom{0}5 & \phantom{0}50 & 200 & \phantom{0}69.03126(38) & 2.285(57) & 44.2(2) \\
           & \phantom{0}6 & \phantom{0}60 & 240 & \phantom{0}69.92349(80) & 1.948(38) & 39.2(2) \\
           & \phantom{0}7 & \phantom{0}80 & 240 & \phantom{0}70.80712(94) & 1.849(26) & 37.7(2) \\
           & \phantom{0}8 & \phantom{0}80 & 280 & \phantom{0}71.64759(94) & 1.807(31) & 36.7(4) \\
          & 10 & \phantom{0}20 & 200 & \phantom{0}73.1522(34)  &           &          \\ 
        & $\infty$& &     &              & 1.619(84) & 33.5(6) \\
        \hline
        16 & \phantom{0}6 & \phantom{0}40 & 160 & 179.8508(20) & 5.336(97) & 
    \end{tabular}
    \caption{Summary of results: critical values $\beta_c$ of the inverse coupling $\beta=2\,N/g_0^2$, interface tension $\sigma$, and latent heat $L$ (both in units of appropriate powers of the critical temperature $T_c$) for different numbers of colors $N$ and lattice size used $N_t\times N^2_{x,y}\times N_z$. 
    The continuum limit extrapolations are labeled by $N_t = \infty$. See Tables~\ref{tab:sigma}-\ref{tab:sigmacontinuumlim} in Appendix~\ref{app:ittables} for more details on the listed results on $\sigma$, and Tables~\ref{tab:latentheatcompdat}-~\ref{tab:lhlargeNlim} in Appendix~\ref{app:lhtables} for more details on the latent heat results.}
    \label{tab:results}
\end{table}
\setlength{\tabcolsep}{0.75em}

\begin{figure*}[tb]
    \begin{minipage}[t]{0.5\linewidth}
    \vspace{0pt}
    \centering
      {\small \qquad $\SU{4}$}\\[-1pt]
      \includegraphics[width=0.95\linewidth]{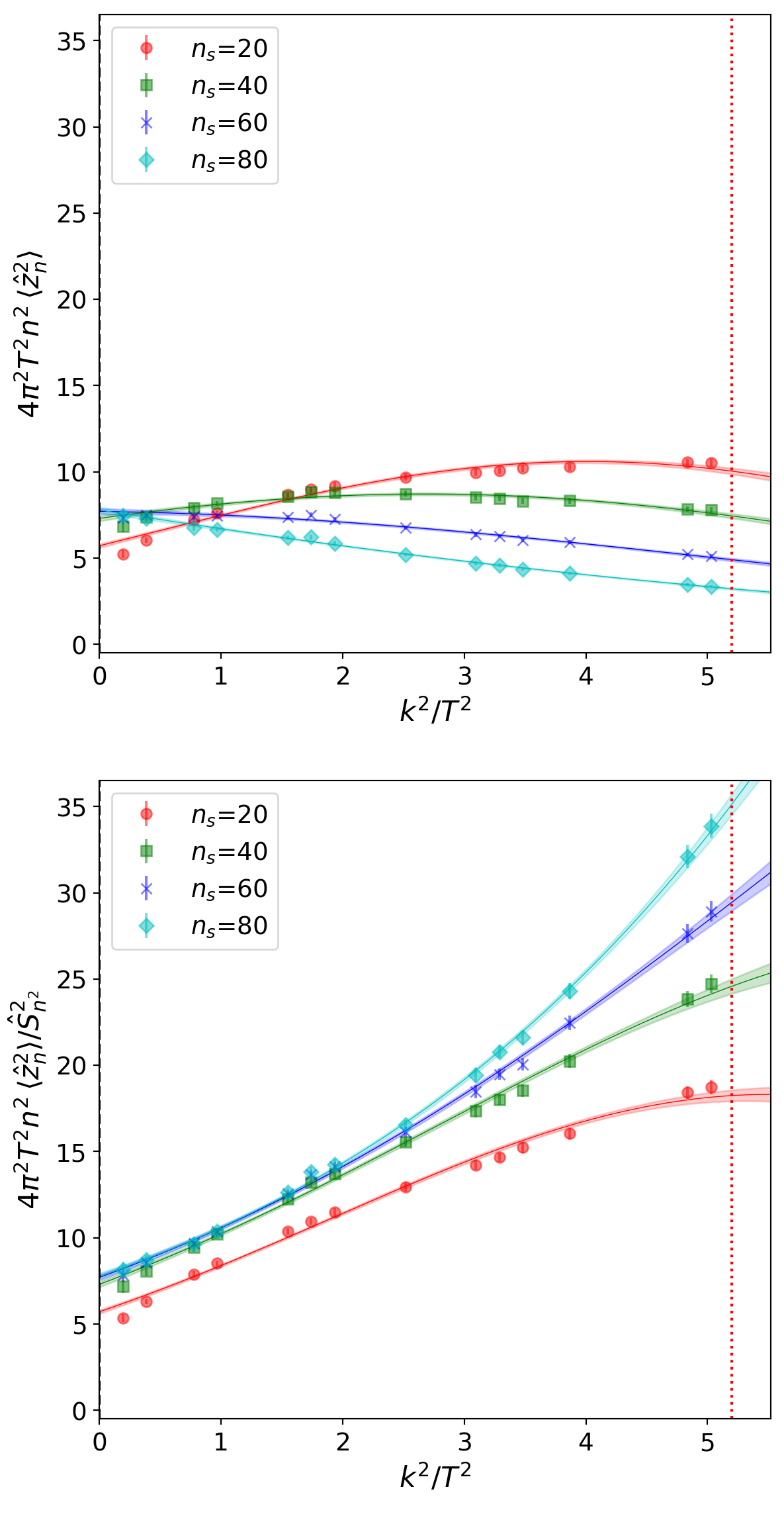}
    \end{minipage}\hfill
    \begin{minipage}[t]{0.5\linewidth}
    \vspace{0pt}
    \centering
      {\small \qquad $\SU{10}$}\\[-1pt]
      \includegraphics[width=0.95\linewidth]{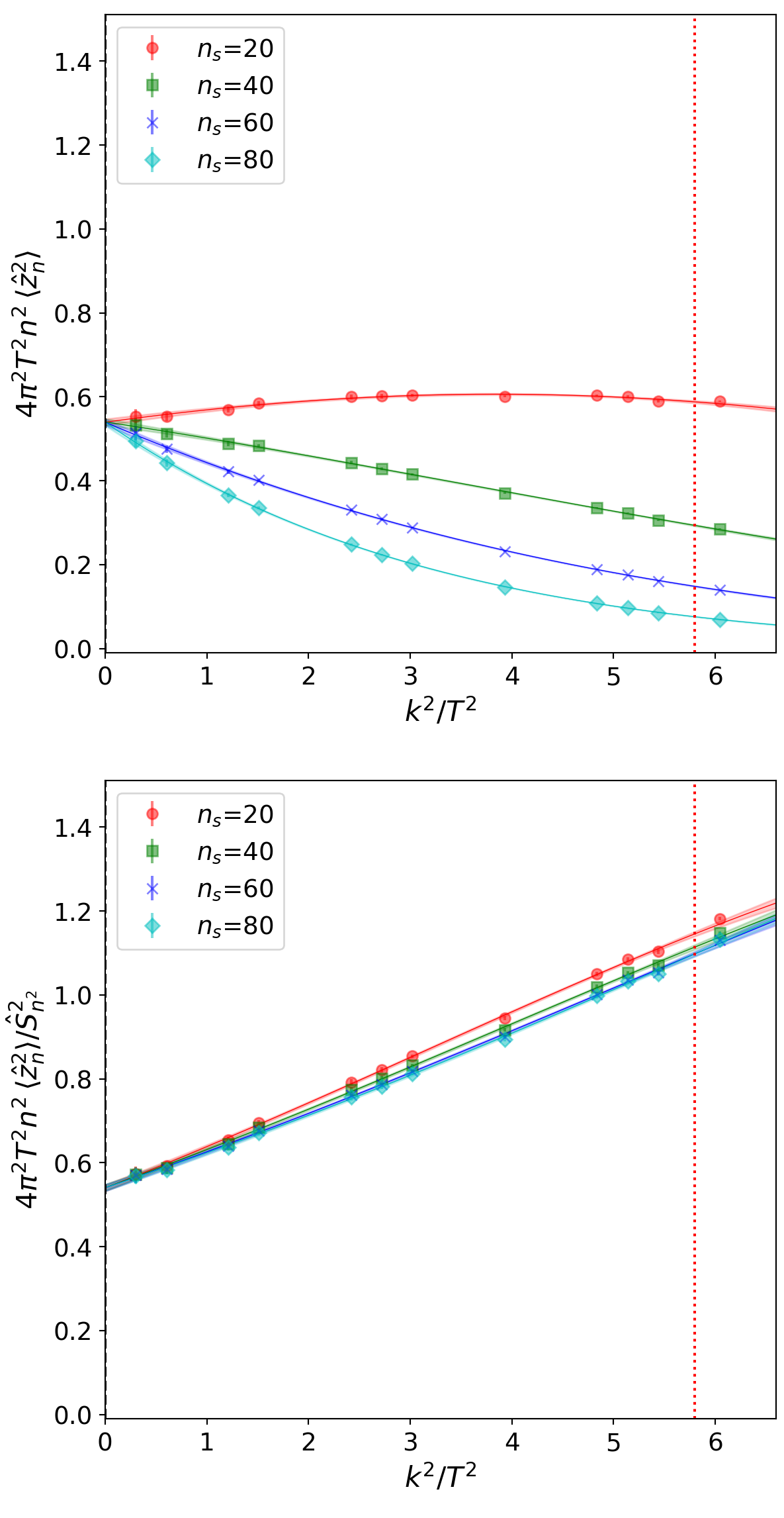}
    \end{minipage}
    \caption{\emph{Top:} extrapolation of the interface Fourier modes to $k\rightarrow 0$ limit at different number of smearing steps for SU(4) at $\beta=10.94345$ on a $100^2 \times 400 \times 7$ lattice (left) and SU(10) at $\beta=70.80712$ on a $80^2\times 240 \times 7$ lattice (right).
    \emph{Bottom:} same after kernel correction. Data collapse at sufficiently large smearing levels. The $k\rightarrow 0$ extrapolation gives the value of $T^3/\sigma$. The continuous curves represent fits of the form Eq.~\eqref{eq:spectrumfit} to the data. The dotted, vertical lines mark the upper limit of the fit range. The fit results are listed in Table~\ref{tab:itzeromodeextrapfitres} of Appendix~\ref{app:ittables}. 
    }
    \label{fig:fouriermodes}
\end{figure*}

Let us now return to the interface Fourier modes, Eq.~\eqref{eq:zhat}. In the upper panels of Fig.~\ref{fig:fouriermodes} we show examples of the behavior of $4\pi^2\,n^2\,T^2 \langle |\hat z(n)|^2\rangle$ at small $k^2 = (2\,\pi\,n/L)^2$ (where $L=N_{x,y}$), measured for a wide range of smearing levels, for SU(4) and SU(10).  As $k^2 \rightarrow 0$, this quantity approaches $T^3/\sigma$ (cf. Eq.~\eqref{eq:zhat}).

As expected, smearing decreases the magnitude of the higher $k$ Fourier modes, but the extrapolation to $k^2\rightarrow 0 $ (detailed in the following Sec.~\ref{ssec:itzeromodeextrapolation}) remains stable for a wide range of smearing levels.  For SU(10) the convergence is faster, all smearing levels between 20 and 80 steps extrapolate to the same value, whereas for SU(4) we need to do 60 to 80 smearing steps to reach convergence. This is obviously due to the fact that SU(4) has smaller surface tension and larger correlation length than SU(10).

In order to make the measurements at different smearing levels comparable, we show in Appendix \ref{app:smearingkernel} how the action of the smearing can be ``undone'', using {\em kernel-corrected} result:
\begin{equation}
    \langle |\hat z^{(n_s)}(n_x,n_y)|^2\rangle_{\text{corr.}} =  
    \frac{\langle |\hat z^{(n_s)}(n_x,n_y)|^2\rangle}{[\hat S(n_x,n_y,0)]^{2\,n_s}}\,.\label{eq:kernelcorrectedamp}
\end{equation}
Here $\hat S$ is the Fourier transform of the smearing kernel Eq.~\eqref{eq:smearingkernel}.  Examples of kernel-corrected results are shown in the bottom panels of Fig.~\ref{fig:fouriermodes}. At sufficiently high smearing levels the low-$k^2$ data are seen to collapse on a single line, demonstrating that the systematics is well under control.

We can also observe that after kernel correction, the measured values of $n^2 \ssavof{\ssabs{\hat z\of{n}}^2}$ increase approximately linearly with $k^2/T^2$.  This behavior can be understood by higher-order effects in interface fluctuations and by the Polyakov line fluctuations in the bulk phases, as discussed in Appendix~\ref{app:fluctuations}.

\begin{figure}[h]
\centering
{\small \qquad $\SU{5}$}\\[-1pt]
\includegraphics[width=0.99\linewidth]{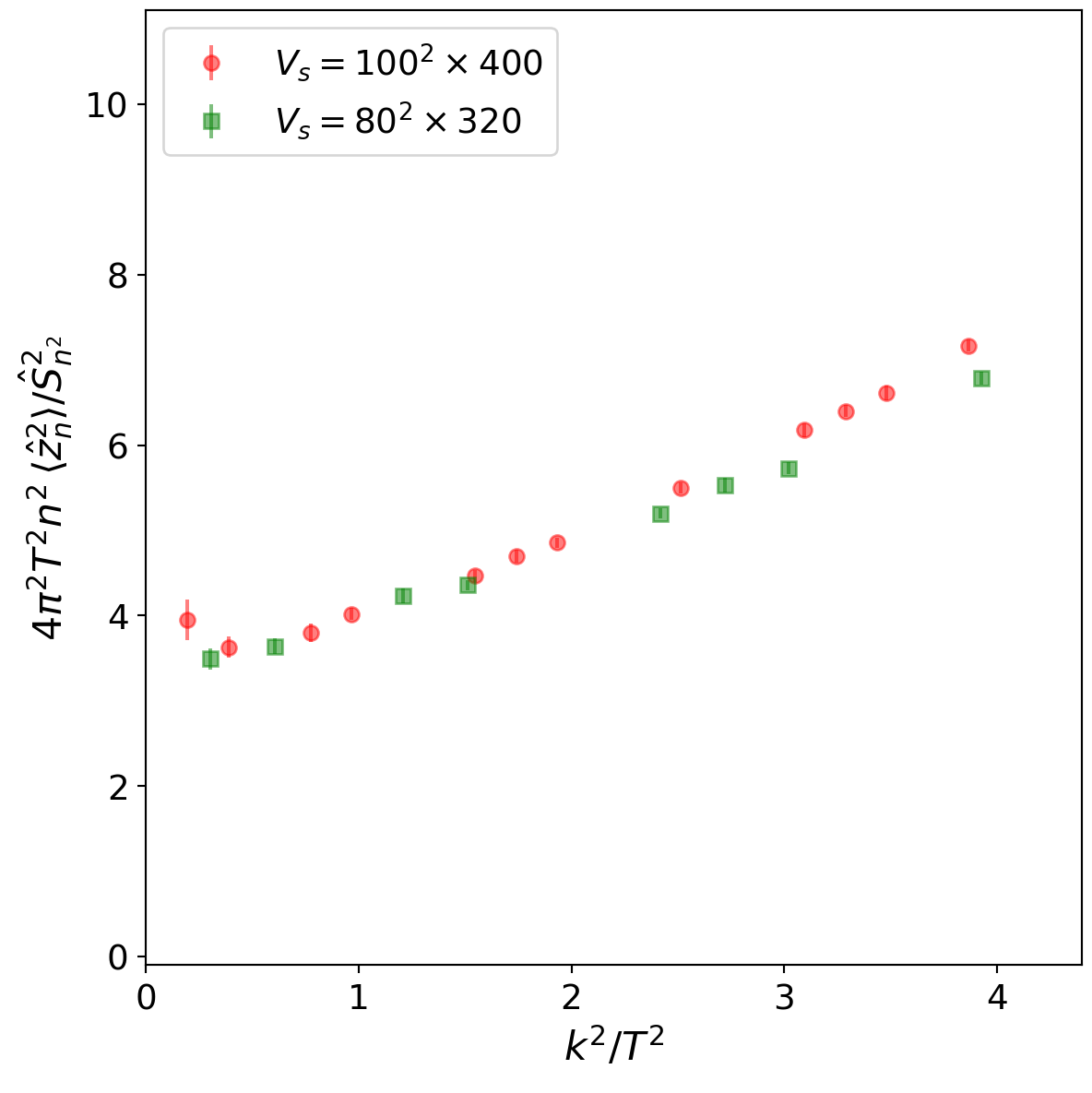}\\[3pt]
{\small \qquad $\SU{10}$}\\[-1pt]
\includegraphics[width=0.99\linewidth]{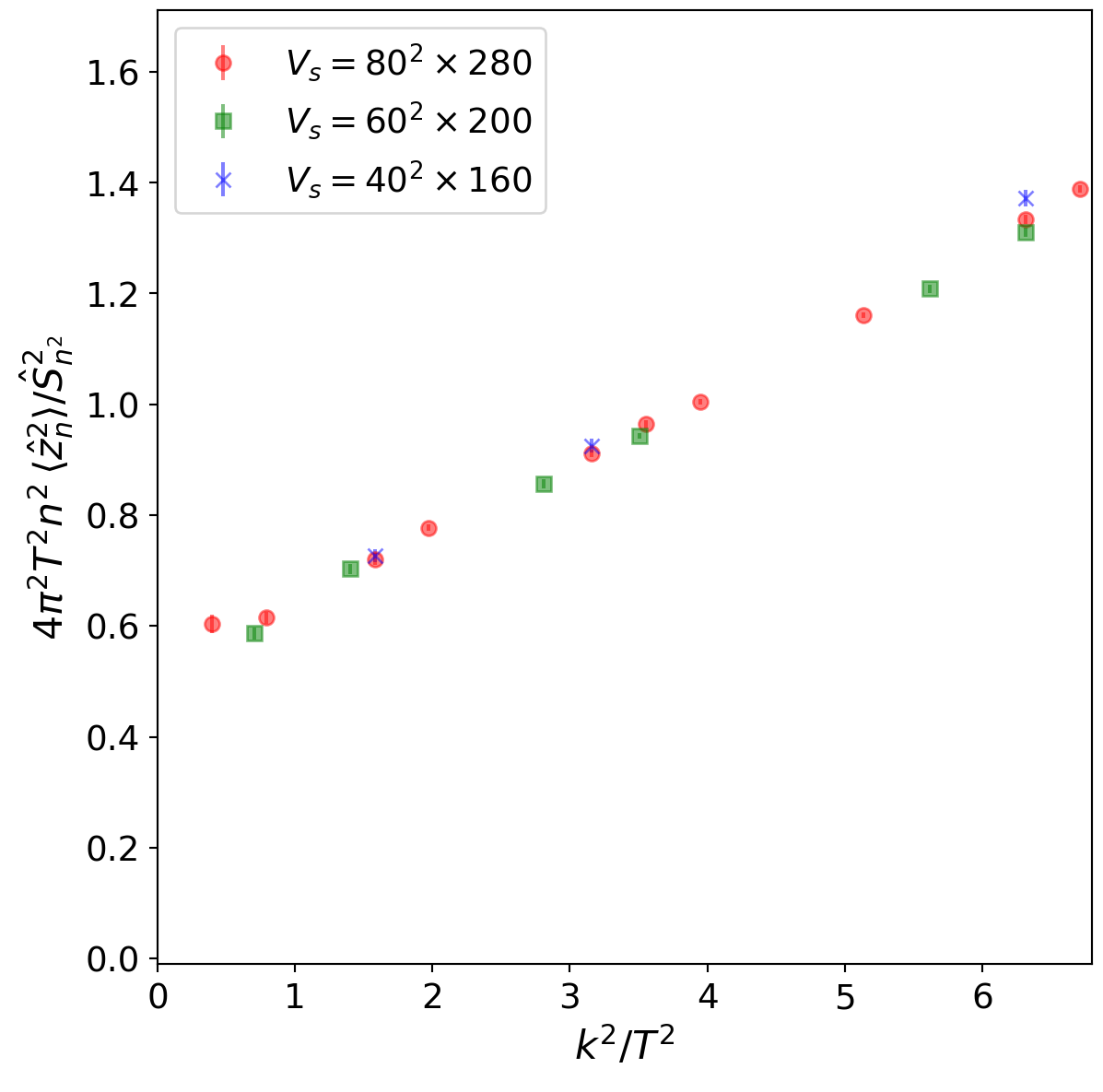}
\caption{Rescaled spectrum after kernel correction as used for extracting $T^3/\sigma$ via zero mode extrapolation (cf. Fig.~\ref{fig:fouriermodes}), plotted for $\SU{5}$ on a lattice of temporal size $N_t=7$ and two different spatial sizes, and for $\SU{10}$ on a lattices of temporal size $N_t=8$ and three different spatial sizes. In both examples the data obtained from different spatial lattices line up reasonably well.}
\label{fig:iffinitevolcomp}
\end{figure}

Our volumes are sufficiently large that the finite volume effects can be neglected
when extrapolating $n^2 \ssavof{\ssabs{\hat z\of{n}}^2}$ to $k^2\to 0$ to extract the inverse interface tension. The latter is illustrated in Fig.~\ref{fig:iffinitevolcomp} where kernel corrected data for $n^2 \ssavof{\ssabs{\hat z\of{n}}^2}$ are shown for $\SU{5}$ (top panel) and $\SU{10}$ (bottom panel) for two resp. three different spatial lattice sizes. In both examples, the data obtained with different volumes align very well, extrapolating to the same result within errors. Thus, we present our results using the largest volumes, all of which had linear sizes of at least 10/$T_c$.

\subsection{Long-wavelength extrapolation}\label{ssec:itzeromodeextrapolation}

We extrapolate the non-kernel-corrected measurements of $k^2\,\langle|\hat z(k)|^2\rangle$ to $k^2 \rightarrow 0 $ by fitting the function
\begin{equation}
    f\of{x} = c_1 \exp(c_2\,x + c_3\,x^2),\label{eq:spectrumfit}
\end{equation}
to the data $\sscof{\ssof{k^2/T_c^2,\ssavof{\ssabs{\hat{z}\of{k}}^2}\,k^2\,T_c^2\,N_{x,y}^2}\,|\,k^2/T_c^2\lsim 6}$, where each data point is weighted by the inverse of its squared error. The restriction $x \equiv k^2/T_c^2 \lsim 6$ implies interface fluctuation wavelengths $\lambda \gsim 2.5/T_c$.  
For the kernel-corrected measurements, we substitute $\ssavof{\ssabs{\hat{z}}^2}$ with $\ssavof{\ssabs{\hat{z}_\text{corr.}}^2}$ from Eq.~\eqref{eq:kernelcorrectedamp}. 

The exponential form of the fit function is motivated by the fact that the non-kernel-corrected measurements decrease exponentially as functions of $k$ at high smearing levels, which was also used in ref.~\cite{Moore:1996bn}.  With the added subleading $\order\ssof{x^2}$ correction in the exponential it allows for good fits in a wider fit range.

The fits to kernel-corrected and non-kernel-corrected data give compatible results, but our final results are presented with the kernel correction applied, since it generally allows for a better quality of the fits. Examples for SU(4) and SU(10) at $N_t=7$ are shown in Fig.~\ref{fig:fouriermodes} and the fit results are listed in Table~\ref{tab:itzeromodeextrapfitres} of Appendix~\ref{app:ittables}. The widths of the error bands are computed from the covariance matrix $\Sigma=\ssof{\sigma^{i j}}_{i,j\in \cof{1,\ldots}}$ of the fitted parameters, using the error propagation formula
\begin{equation}
    \delta f\of{x} = \sqrt{\sum\limits_{i,j} \sigma^{i j} \frac{\partial f\of{x}}{\partial c_i}\,\frac{\partial f\of{x}}{\partial c_j}}\label{eq:errorbands}\ ,
\end{equation}
with $f(x)$ given by Eq.~\eqref{eq:spectrumfit}.
%
From the data shown in Table~\ref{tab:itzeromodeextrapfitres} of Appendix~\ref{app:ittables}, we can observe that the results obtained from fits to corresponding kernel-corrected and non-kernel-corrected data differ predominantly by the fitted values for the parameter $c_2$, while the $k^2\rightarrow 0$ amplitude, $c_1$, and the \refcorrb[subleading]{higher order} correction parameter, $c_3$, are found to be the same for equal numbers of smearing steps $n_s$. We also see that $c_2$ varies signficantly between smearing levels in non-kernel-corrected fits, while it converges quickly with increasing numbers of smearing steps in the kernel-corrected case.

A summary of the data and the corresponding extrapolation results for each SU($N$) and $N_t$ used for further analysis is given in Table~\ref{tab:sigma} of Appendix~\ref{app:ittables}. 

\subsection{Continuum extrapolation and large-$N$ limit}\label{ssec:itcontinuumextrapolation}
\begin{figure}[h]
\centering
\includegraphics[width=0.99\linewidth]{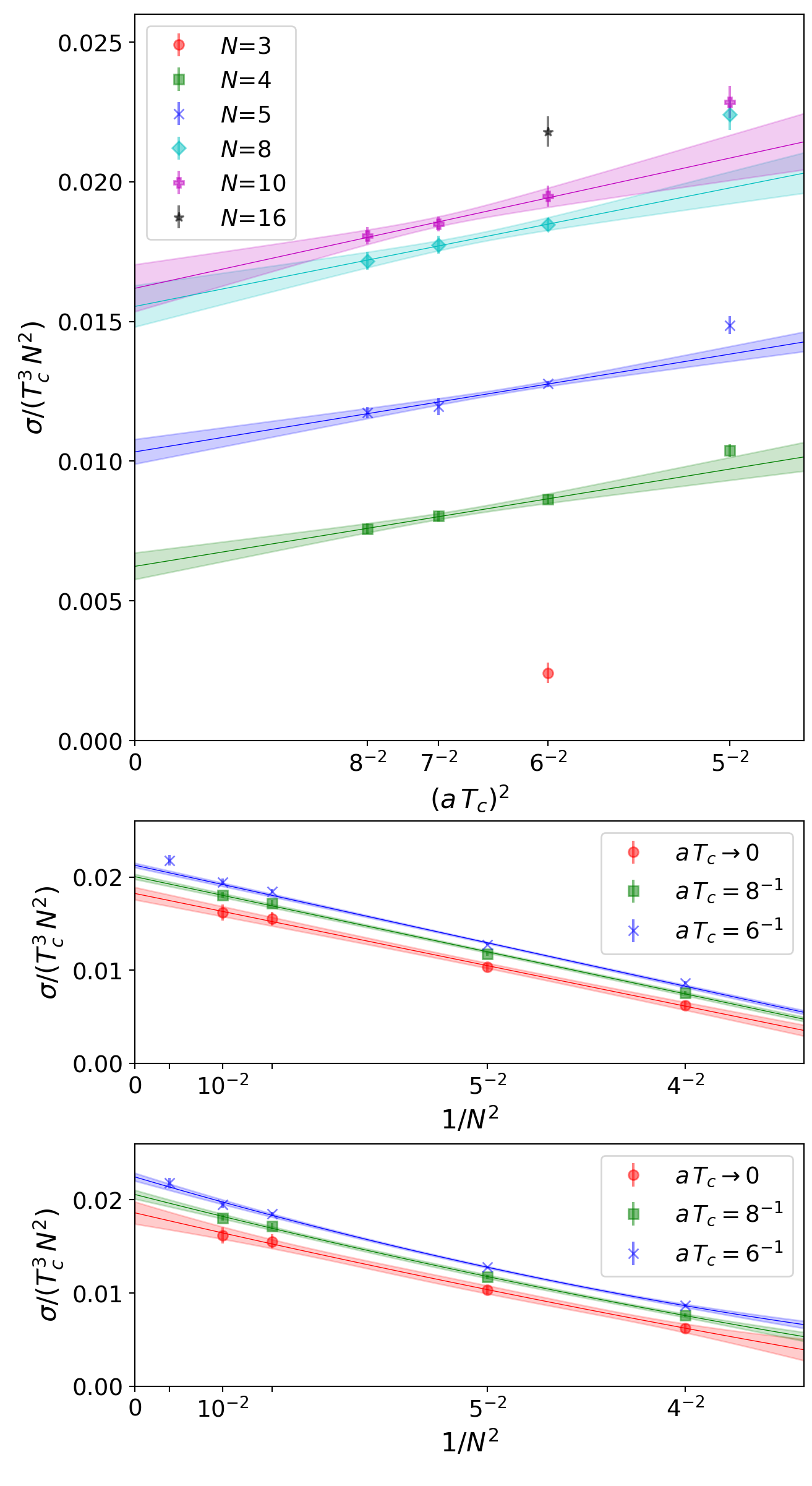}
\caption{\emph{Top:} Continuum extrapolation of the interface tension using the data for $N=4\ldots 10$ and $N_t=6 \ldots 8$ from Table~\ref{tab:sigma} of Appendix~\ref{app:ittables}. For comparison, we also include SU(3), $N_t=6$ value \cite{Iwasaki:1993qu}. \emph{Middle:} large-$N$ extrapolation of the obtained continuum results for the interface tension (red) (cf. Table~\ref{tab:sigmacontinuumlim} \refcorrb{of Appendix~\ref{app:ittables}}) as well as of the finite lattice spacing data corresponding to $a\,T_c=1/N_t$ with $N_t=6$ (blue) and $N_t=8$ (green), using the linear ansatz from Eq.~\eqref{eq:itlargeNfit}. \emph{Bottom:} same as middle panel, but with the quadratic ansatz \eqref{eq:itlargeNfitquad}. \refcorrb{The large-$N$ fit results are listed in Table~\ref{tab:sigmalargeNlim} of Appendix~\ref{app:ittables}.}}
\label{fig:iftensionfits}
\end{figure}

The continuum limit is obtained for each $\SU{N}$ group by fitting the linear function, 
\begin{equation}
f\of{x}=c_1 + c_2\,x \text{~~with~} x = \of{a\,T_c}^{2} = 1/N_t^2,\label{eq:itcontinuumfit}
\end{equation}
to the data $\sscof{\ssof{1/N_t^2,\sigma/\ssof{T_c^3\,N^2}}\,|\, N_t>5}$, obtained from the fit results for $T_c^3/\sigma$ from Table~\ref{tab:sigma} of Appendix~\ref{app:ittables}. The fitting is performed with each data point being weighted by the inverse of its squared error. Note that we exclude $N_t=5$ from the continuum extrapolation; including it makes the fits markedly worse even with subleading corrections. The fits are shown in the upper panel of Fig.~\ref{fig:iftensionfits}, with error bands computed using Eq.~\eqref{eq:errorbands} applied to Eq.~\eqref{eq:itcontinuumfit}. The fit results are summarized in Table~\ref{tab:sigmacontinuumlim} of Appendix~\ref{app:ittables}.

We finally perform the large-$N$ extrapolation of our interface tension data, by fitting the linear function
\begin{equation}
f\of{x}=c_1 + c_2\,x \text{~~with~} x = 1/N^2 ,\label{eq:itlargeNfit}
\end{equation}
to the data set $\sscof{\ssof{1/N^2,\sigma/\ssof{T_c^3\,N^2}}\,|\,N\in\fof{4,10}}$ given by the continuum results from Table~\ref{tab:sigmacontinuumlim} of Appendix~\ref{app:ittables}. Again, data points are weighted by the inverse of their squared errors. The large-$N$ behavior of the continuum interface tension is well described by the leading $N^2$-dependency:
\begin{equation}
\frac{\sigma}{T^3} = 0.0182(7) N^2 - 0.194(15)\ ,\label{eq:itlargeNlinfitres}
\end{equation}
with $\chi^2/\text{\#dof} = 0.2$ (cf. red band in middle panel of Fig.~\ref{fig:iftensionfits}).

For comparison, we also extract large-$N$ limits of two finite lattice spacing data sets, corresponding to $a\,T_c=1/N_t$ with $N_t=6,8$ (blue and green bands in middle panel of Fig.~\ref{fig:iftensionfits}). It turns out that in the latter cases, the linear ansatz, Eq.~\eqref{eq:itlargeNfit}, does not work particularly well, as illustrated by the relatively large $\chi^2/\text{\#dof}$ values in Table~\ref{tab:sigmalargeNlim} of Appendix~\ref{app:ittables} for these data sets. We therefore repeated the fits with a quadratic anstz, 
\begin{equation}
f\of{x}=c_1 + c_2\,x + c_3\,x^2 \text{~~with~} x = 1/N^2 .\label{eq:itlargeNfitquad}
\end{equation}
Since for the continuum data, the linear ansatz performed already well, the higher order correction is in this case not well constrained and the overall uncertainty in the fit increases slightly compared to the linear ansatz. The quadratic fit result reads,
\begin{equation}
\frac{\sigma}{T^3} = 0.0186(12) N^2 - 0.220(71) + 0.35(92)/N^2\ ,\label{eq:itlargeNquadfitres}
\end{equation}
with $\chi^2/\text{\#dof} = 0.25$ (cf. red band in bottom panel of Fig.~\ref{fig:iftensionfits}), which is fully compatible with the linear result Eq.~\eqref{eq:itlargeNlinfitres}. For the finite lattice spacing data, however, the quadratic ansatz \refcorrb[\eqref{eq:itlargeNquadfitres}]{\eqref{eq:itlargeNfitquad}} results in significantly better qualities of fit, as the $\chi^2/\text{\#dof}$ for these fits in Table~\ref{tab:sigmalargeNlim} show. We note that for the $N_t=6$ case, the $N=16$ data point aligns well with the quadratic fit result. Inclusion of this data point in the quadratic fit would change the fit result only marginally and well within the error bounds of the current fit results stated in Table~\ref{tab:sigmalargeNlim}.

\refcorrb{We conclude this section by noting that our continuum data for $\sigma/T^3$ as function of $N$ cannot be well described by a linear leading large $N$ behavior: fitting
\begin{equation}
f\of{x}=c_1 + c_2\,x \text{~~with~} x = 1/N ,\label{eq:itlargeNfitlinN}
\end{equation}
to the data set $\sscof{\ssof{1/N,\sigma/\ssof{T_c^3\,N}}\,|\,N\in\fof{4,10}}$, obtained from the continuum results in Table~\ref{tab:sigmacontinuumlim} of Appendix~\ref{app:ittables}, leads to
\begin{equation}
\frac{\sigma}{T^3} = 0.213(8) N - 0.77(4)\ ,\label{eq:itlargeNlinfitreslin}
\end{equation}
with $\chi^2/\text{\#dof} = 13.7$.}

\section{Latent heat}\label{sec:latenheat}

The latent heat can be calculated from the discontinuity of the derivative of the free energy density
\begin{equation}
    L = \Delta \frac{\mathrm{d}}{\mathrm{d}T}\frac F V = \frac {T_c^2}{V} \Delta \frac{\mathrm{d}}{\mathrm{d}T} \ln Z 
\end{equation}
where $\Delta$ is the discontinuity at $T=T_c$.  On the lattice this can be expressed in terms of the
discontinuity of the plaquette expectation value $\Delta \langle U_\Box \rangle$ at $\beta = \beta_c$:
\begin{equation}
    \frac L{T_c^4} = -6\,N_t^4 { \frac{\mathrm{d} \beta}{\mathrm{d} \ln a} } {\Delta \langle U_\Box \rangle}.
    \label{eq:latent}
\end{equation}

\subsection{Lattice beta-function}\label{ssec:latbetafunc}
The lattice $\beta$-function ${\mathrm{d} \beta}/{\mathrm{d} \ln a}$ is often evaluated using string tension or gradient flow for scale setting.  However, since we have accurate measurements of $\beta_c$ as functions of $N_t = 1/(a T_c)$, we use $T_c$ as the fixed physical scale. 

Because the data is discrete, we interpolate the measurements to continuous $N_t = 1/(aT_c)$ to evaluate the derivative, using a generalization of the fit ansatz from~\cite{Lucini:2005vg} to fit the data for all $N$ simultaneously:
\begin{multline}
\log\of{a\,T_c}=-\log\of{N_t}=\\
f\of{x}=-\log\of{N_{t,0}}-\frac{12\pi^2}{11}\of{x-x_0\of{N}}\\
+\sum_{i=1}^{n_{\text{ord}}}\sum_{j=0}^{n_{c,\text{ord}}}c_{i,j}\of{\frac{1}{x}-\frac{1}{x_0\of{N}}}^{i}\frac{1}{N^j}\ ,\label{eq:interpg}
\end{multline}
where $x=\beta/N^2$, $x_0\of{N}=\beta_c\of{N_{t,0},N}/N^2$, and the expansion orders are set to $n_{\text{ord}}=4$, $n_{c,\text{ord}}=2$. 
We use $N_{t,0}=7$, which is approximately in the middle of the $N_t$ domain for which we have data, but the final results are not sensitive to this choice. For this choice of $N_{t,0}$ the function $x_0\of{N}$ in the range $N\in\fof{4,10}$ is well described by the following interpolation to our data:
\begin{multline}
x_0\of{N}=0.71329806(1)+0.00198546(4)/N\\
-0.58432233(22)/N^2+0.42305467(43)/N^3\ .
\end{multline}

It turns out that the derivative $\mathrm d \beta/\mathrm d\ln a$ is sensitive to the details of the parametrization of the fit function at the ends of the interpolated range. This is especially significant at $N_t=8$, corresponding to our smallest production lattice spacing, which has an effect on the continuum extrapolation.  To stabilize the fits we added a set of relatively small volume simulations at $N_t = 10$ for all $N$.  This gives us lower-precision results for $\beta_c$ at $N_t=10$, as shown in Table \ref{tab:results}. Nevertheless, these values are sufficient to anchor the value of the derivative at $N_t=8$, and we do not evaluate the derivatives at $N_t=10$.  The parameters given in Eq.~\eqref{eq:interpg} include the $N_t=10$ results.

Since the measurement errors are in $x=\beta/N^2$ instead of $\log\of{N_t}$, we initially perform an unweighted fit and then convert the $x$ errors into effective errors for the corresponding values of $\log\of{N_t\of{N}}$, using the local derivatives of the fitted function. The effective errors for $\log\of{N_t\of{N}}$ are then used to iteratively perform weighted fits, until the fit result and the effective errors it produces for the values of $\log\of{N_t\of{N}}$ stop changing. The final fit obtained with this procedure yields $\chi^2/\text{\#dof}=0.3$. 

\begin{figure}[t]
\centering
\includegraphics[width=0.99\linewidth]{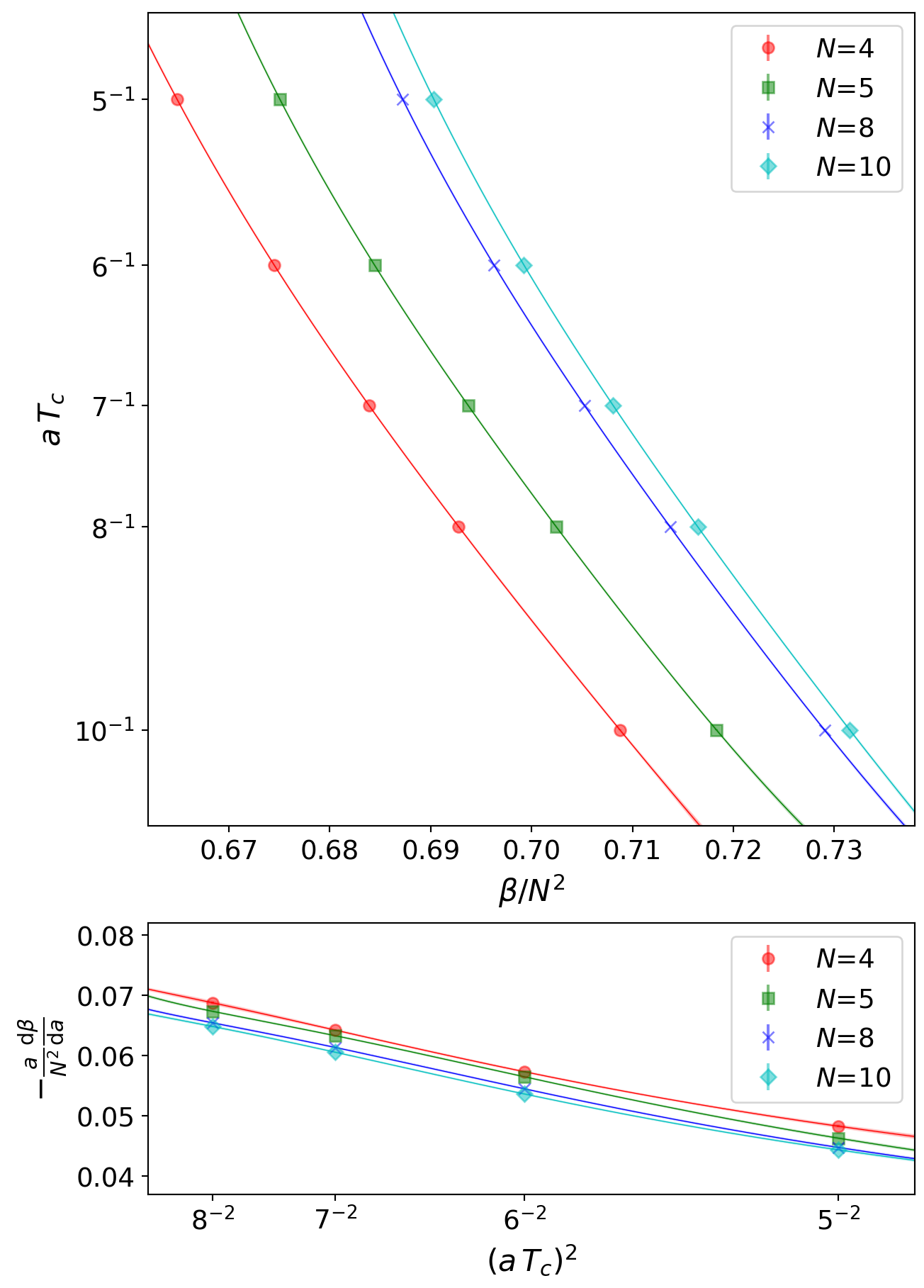}
\caption{\textit{Top:} running of the lattice spacing as function of $\beta/N^2$, obtained form the data for $\beta_c\of{N,N_t}$ from Table~\ref{tab:results} for $N=4,5,8,10$, $N_t=5,6,7,8,10$, and the global fitting procedure described below Eq.~\eqref{eq:interpg}. \textit{Bottom:} lattice beta function obtained from the above global fit, plotted for $N=4,5,8,10$. The plot markers indicate the values relevant for the latent heat computations, listed in fourth column of Table~\ref{tab:latentheatcompdat} of Appendix~\ref{app:lhtables}.}
\label{fig:fitsforbetafunc}
\end{figure}

The fitted values for the parameters $c_{i,j}$ for $i=1,\ldots,n_{\text{ord}}$, $j=0,\ldots,n_{c,\text{ord}}$ are listed in Table~\ref{tab:betacglobalfitres} of Appendix~\ref{app:lhtables} and the function is plotted in the upper panel of Fig.~\ref{fig:fitsforbetafunc} for $N=4,5,8,10$, together with the fitted data points. 

The lower panel of Fig.~\ref{fig:fitsforbetafunc} shows the lattice beta function (divided by $N^2$) as function of $\of{a\,T_c}$, obtained from the fitted function. The plot markers in the latter represent the values relevant for the latent heat determination, which are listed in the 3rd column of Table~\ref{tab:latentheatcompdat} of Appendix~\ref{app:lhtables}.

\subsection{Plaquette discontinuity}\label{ssec:plaquettediscontinuity}

\begin{figure}[t]
    \centering
\includegraphics[width=0.95\columnwidth]{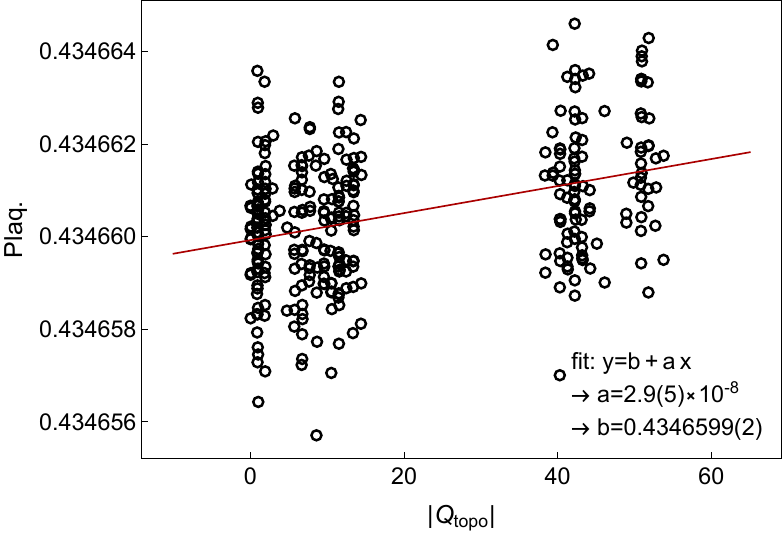}
\caption{Dependency of average Wilson plaquette action on the topological charge for $\SU{8}$ lattice gauge theory in the confined phase, simulated on a $64^3\times 8$ lattice at the pseudo-critical inverse gauge coupling $\beta_{\mathrm{pc}}\approx 45.677$. The data is obtained from 8 independent Markov chains that were initialized to start in different topological sectors. Topological charge measurements were performed every 5000 sweeps, with a sweep amounting to 1 heatbath and 5 overrelaxation updates. The plaquette values assigned to each topological charge measurement are averages of the plaquette measurements carried out between two subsequent topological charge measurements. This is reasonable since the topological charge evolves only very slowly in these systems.}
\label{fig:plaqvstopo}
\end{figure}

We measure the plaquette discontinuity by separate simulations at $\beta_c$, prepared so that the system is fully in the confined or the deconfined phase, and using large enough volumes so that the system remains in the same phase throughout the simulation. The measured plaquette values for confined and deconfined phases are listed in the 5th resp. 6th comlumn of Table~\ref{tab:latentheatcompdat}. Due to the large volumes, the confined phase simulations suffer from topological freezing. Since the Euclidean gauge action grows with the absolute value of the topological charge (cf. Fig.~\ref{fig:plaqvstopo}), we incorporated a systematic uncertainty of order $2\times 10^{-6}$ in the error of the plaquette difference in the second to last column of Table~\ref{tab:latentheatcompdat} of Appendix~\ref{app:lhtables}. 

Combining our lattice beta function and plaquette discontinuity results yields for each $N=4,5,8,10$ and $N_t=5,6,7,8$ the corresponding latent heat values listed in the last column of Table~\ref{tab:latentheatcompdat} of Appendix~\ref{app:lhtables}. 

\subsection{Continuum extrapolation and large-$N$ limit}\label{ssec:lhcontinuumextrapolation}
The continuum extrapolated values ($\sim\,N_t=\infty$) for the latent heat are obtained analogously to those for the interface tension, i.e., by fitting for each $\SU{N}$ the linear function Eq.~\eqref{eq:itcontinuumfit} to the data sets $\sscof{\ssof{1/N_t^2,L/\ssof{N^2\,T_c^4}}\,|\, N_t>5}$ where we again exclude $N_t=5$. The fits are shown in the upper panel of Fig.~\ref{fig:latentheatfits} and the fit results are listed in Table~\ref{tab:latheatcontinuumlim} of Appendix~\ref{app:lhtables}.  

\begin{figure}[t]
\centering
\includegraphics[width=0.99\linewidth]{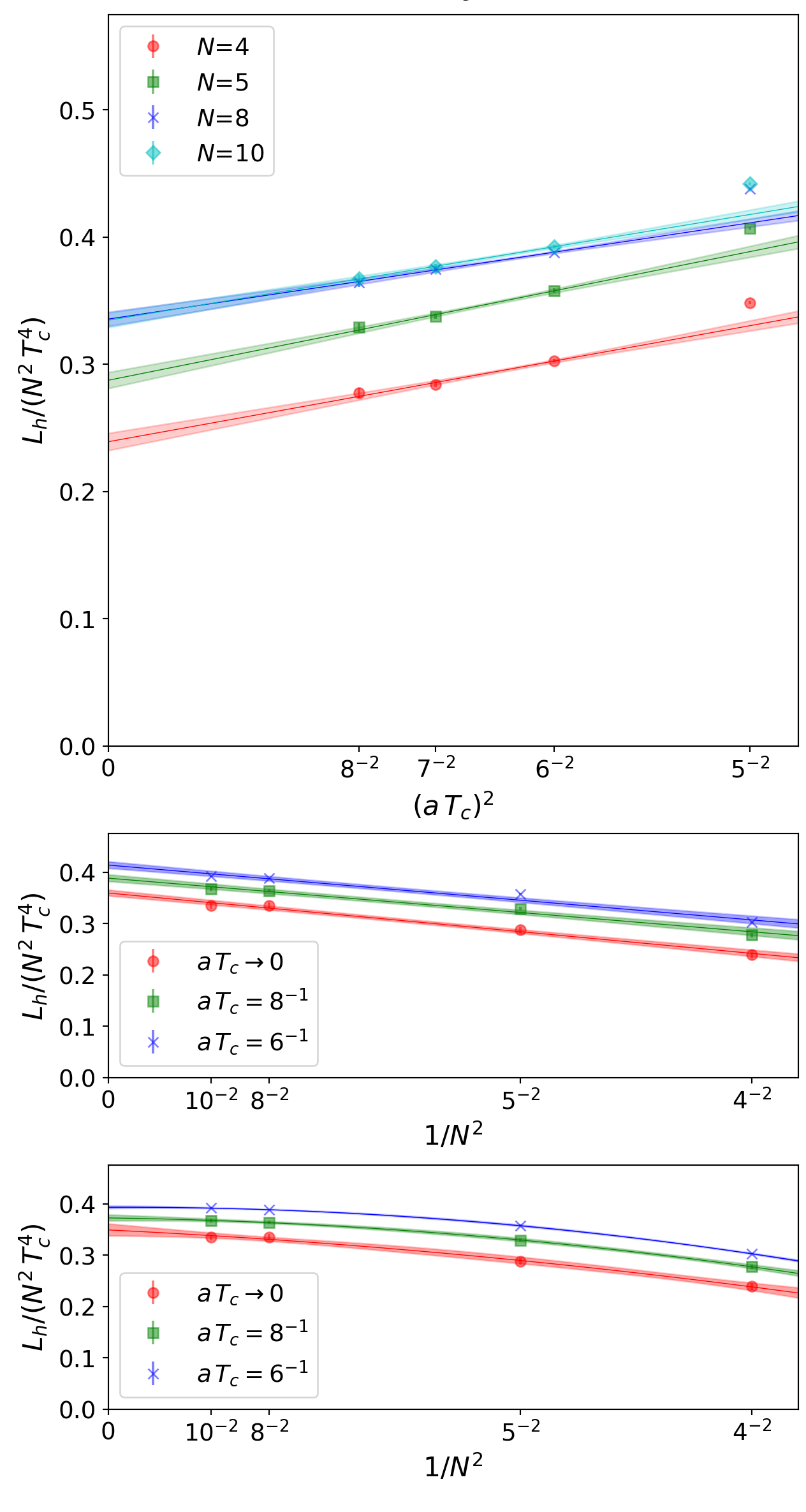}
\caption{\textit{Top:} Latent heat data and continuum extrapolation fits obtained with the linear ansatz from Eq.~\eqref{eq:itcontinuumfit} on the data for $N_t=6,7,8$. Continuum fit results are listed in Table~\ref{tab:latheatcontinuumlim} of Appendix~\ref{app:lhtables}. \textit{Middle:} Large $N$ extrapolation of the latent heat continuum (red) and finite lattice spacing values (green, blue), using the linear ansatz from \refcorrb[\eqref{eq:itlargeNlinfitres}]{Eq.~\eqref{eq:itlargeNfit}} on the data for $N=4,5,8,10$. \textit{Bottom:} Same as middle panel but using the quadratic ansatz from Eq.~\refcorrb[\eqref{eq:itlargeNquadfitres}]{\eqref{eq:itlargeNfitquad}} on the data for $N=4,5,8,10$. Large $N$ extrapolation fit results can be found in Table~\ref{tab:lhlargeNlim} of Appendix~\ref{app:lhtables}.}
\label{fig:latentheatfits}
\end{figure}

Also for the large-$N$ extrapolation we proceed in the same way as with the interface tension: we fit the linear ansatz \eqref{eq:itlargeNfit} to the data set $\sscof{\ssof{1/N^2,L/\ssof{N^2\,T_c^4}}\,|\, N\in\fof{4,10}}$, using the continuum and finite lattice spacing results from Table~\ref{tab:latheatcontinuumlim} of Appendix~\ref{app:lhtables}. The fit results are reported in Table~\ref{tab:lhlargeNlim} of Appendix~\ref{app:lhtables}. We find that the large-$N$ behavior of the continuum latent heat is reasonably described by a leading+subleading $N^2$-dependency:
\begin{equation}
\frac{L}{T^4} = 0.360(6) N^2 - 1.88(17)\ ,\label{eq:lhlargeNlinfitres}
\end{equation}
with $\chi^2/\text{\#dof} = 1.26$ (cf. red band in middle panel of Fig.~\ref{fig:latentheatfits}). 

For the finite lattice spacing data corresponding to $N_t=6,8$ (blue and green bands in the middle panel of Fig.~\ref{fig:latentheatfits}) the linear fit ansatz does not work particularly well, resulting in large $\chi^2/\text{\#dof}$ values. 

With the quadratic ansatz \eqref{eq:itlargeNfitquad}, the continuum latent heat is described by
\begin{equation}
\frac{L}{T^4} = 0.349(12) N^2 - 0.97(92) - 12.8(128)/N^2\ ,\label{eq:lhlargeNquadfitres}
\end{equation}
with $\chi^2/\text{\#dof} = 1.25$ (cf. red band in bottom panel of Fig.~\ref{fig:latentheatfits}). The leading behavior is compatible with the linear result Eq.~\eqref{eq:lhlargeNlinfitres}, while the subleading terms are essentially indeterminate. For the finite lattice spacing data, the quadratic ansatz \refcorrb[\eqref{eq:itlargeNquadfitres}]{\eqref{eq:itlargeNfitquad}} yields significantly better qualities of fit, as indicated by the $\chi^2/\text{\#dof}$ values listed in Table~\ref{tab:lhlargeNlim} of Appendix~\ref{app:lhtables}. However, the parameters $c_2$ remain also in these fits essentially indeterminate.

We note here that the latent heat has been determined in SU(3) gauge theory to high precision \cite{Giusti:2025fxu}, with the result $L/T_c^4 = 1.175(10)$.  Our fit Eq.~\eqref{eq:lhlargeNlinfitres} extrapolated to $N=3$ gives $(L/T_c^4)_\textrm{fit} = 1.36(17)$ which is in practice compatible with this result.

\refcorrb{As was the case for the interface tension, also our continuum data for $L/T^4$ as function of $N$ is incompatible with a linear leading large $N$ dependency: fitting the ansatz \eqref{eq:itlargeNfitlinN} to the data set $\sscof{\ssof{1/N,L/\ssof{T_c^4\,N}}\,|\,N\in\fof{4,10}}$, obtained from the continuum results listed in Table~\ref{tab:latheatcontinuumlim} of Appendix~\ref{app:lhtables}, leads to
\begin{equation}
\frac{L}{T^4} = 4.6(4) N - 14.9(19)\ ,\label{eq:lhlargeNlinfitreslin}
\end{equation}
with $\chi^2/\text{\#dof} = 30.3$.}

\section{Conclusions}\label{sec:conclusions}

We have measured the latent heat and the interface tension in the confinement-deconfinement phase transition in SU($N$) gauge theory  with $N=4,5,8$ and $10$. We extracted the large-$N$ limit of both of these quantities,
with the result $\sigma/T_c^3 = 0.0182(7) N^2 - 0.194(15)$ for the interface tension and $L/T_c^4 = 0.360(6) N^2 - 1.88(17)$ for the latent heat, valid for $N \ge 4$. These represent \refcorr[]{a} significant increase in the precision of the results, and indeed, in previous results the $N^2$ \refcorr[behaviour]{behavior} of the latent heat could not be determined unambiguously \cite{Lucini:2005vg}.  

We have also determined the lattice $\beta$-function
in SU($N$) theory using the critical temperature to set the scale, and obtained the critical coupling for $N_t = 6$, $7$, $8$ and $10$.
We used the mixed phase ensemble together with the capillary wave theory to determine the critical coupling and the interface tension. The method avoids the supercritical slowing down which affects the widely used histogram method.


\section{Acknowledgments}\label{sec:acknowledgments}
We acknowledge discussions with Niko Jokela, Guy D. Moore and Marco Panero. T.R. acknowledges support from the Swiss National Science Foundation (SNSF) through the grant no.~210064. T.R., K.R., and A.S. acknowledge support form the European Research Council grant 101142449 and Research Council of Finland grant 354572. The authors thank CSC - IT Center for Science, Finland, for considerable computational resources.

\FloatBarrier

\appendix

\begin{widetext}

\section{Surface tension from the capillary wave theory}
\label{app:fluctuations}

In this appendix we derive the equation \eqref{eq:zhat} relating surface tension $\sigma$ and the amplitude of the fluctuations, and discuss subleading corrections.  We mostly follow the discussion in ref.~\cite{Moore:1996bn}.

Let us assume we have a thin surface of size $L^2$ aligned along $(x,y)$-plane and let $z(x,y)$ denote the surface height.  We also assume periodic boundary conditions to $x$ and $y$ -directions.
At finite temperature $T$ the surface fluctuations are described by the partition function
\begin{equation}
    Z = \int \prod_{x,y} \mathrm d z(x,y)\,e^{-H/T}\,,
    \label{eq:surfaceZ}
\end{equation}
where the surface energy is surface tension $\sigma$ times the area:
\begin{align}
H &= \sigma \int_0^L \mathrm d x\,\mathrm d y\, \sqrt{1 + |\nabla z|^2}
   = \sigma \int_0^L \mathrm d x\,\mathrm d y\, \left[1 + \frac 12 |\nabla z|^2
     - \frac 1 8 |\nabla z|^4 + \ldots \right].  \label{eq:effH}
\end{align}
This defines two-dimensional field theory, where higher derivative terms are unrenormalizable. Thus, at long distances, far from the UV cutoff, their effect vanishes and we are left with a massless free theory, capillary wave theory (CWT) \cite{RowlinsonWidom}. Truncating Eq.~\eqref{eq:effH} at second order derivative, the partition function can be solved with a Fourier transform to momentum space:
\begin{eqnarray}
    \hat z(n_x,n_y) &=& \frac 1 {L^2} \int_0^L \mathrm d x \mathrm d y \, z(x,y) e^{2\pi i (n_x x + n_y y)/L} \label{eq:fmode1} \\
    z(x,y) &=& \sum_{n_x,n_y} \hat z(n_x,n_y) e^{-2\pi i (n_x x + n_y y)/L}\, .
\end{eqnarray}
The partition function then factorizes into a product of zero-centered, one-dimensional Gaussian integrals:
\begin{equation}
Z\propto\prod_{n_x,n_y}\int\dd{\hat{z}\of{n_x,n_y}}\,\exp\of{-\frac{4\,\pi^2\,n^2\,\sigma\,\ssabs{\hat{z}\of{n_x,n_y}}^2}{2\,T}}\ ,
\end{equation}
revealing that
\begin{equation}
 \langle |\hat z(n_x,n_y)|^2\rangle =
\frac{T}{4\,\pi^2\,n^2\,\sigma}\ ,  ~~~ n^2 \equiv n_x^2 + n_y^2 \ne 0\ .
\label{eq:zbar}
\end{equation}
Due to the non-renormalizable operators Eq.~\eqref{eq:zbar} becomes exact only in the limit 
$n^2 \rightarrow 0$, with corrections expected to vanish as $O(n^2/L^2)$.  The negative sign of the first correction term in Eq.~\eqref{eq:effH} implies that the energy of the large $n$ (large $k$) modes at a given amplitude is reduced. Hence the amplitude of the surface fluctuations of these modes can be expected to be larger than given by Eq.~\eqref{eq:zbar}. This is consistent with the linear growth of the kernel corrected $n^2 \langle |\hat z(n)|^2 \rangle$ measurements when plotted against $k^2/T^2$ as in Fig.~\ref{fig:fouriermodes}.

In realistic field theory simulations there are also fluctuations of the order parameter in the bulk phases.  These are not directly associated with the phase interface, but bulk fluctuations near the interface will give rise to deviations in the measurements of the interface location $z(x,y)$. These are not described by the partition function in Eq.~\eqref{eq:surfaceZ}. The length scale of the bulk fluctuations is the bulk correlation length $\xi \ll L$. 
Let us consider long-wavelength modes of $\hat z(n)$, with wavelength $\lambda = 2\pi/|k| = L/|n| \gg \xi$.  
With a surface of area $L^2$, we can estimate that there are $n_f = L^2/\xi^2$ fluctuation domains which overlap with the surface. Assuming the fluctuation is approximately constant within the domain, the integral Eq.~\eqref{eq:fmode1} over the area within one domain gives contribution $\propto \xi^2/L^2 $.
Each domain gives a random deviation to integral Eq.~\eqref{eq:fmode1} to positive or negative direction, so that the total effect is $\propto \sqrt{n_f} \xi^2/L^2$.
Thus, we can estimate that the overall effect of the bulk fluctuations on long-wavelength modes is
\begin{equation}
    n^2 \langle |\hat z(n)|^2 \rangle_\text{bulk fluct.} \propto \frac{n^2\xi^2}{L^2},
\end{equation}
again vanishing linearly in $n^2/L^2 \sim k^2$.  The magnitude of this effect is reduced rapidly as the smearing level $n_s$ is increased, proportional to $\exp[-2\pi\alpha_\rho n_s a^2/\xi^2]$, see Eq.~\eqref{eq:plmodesmearingeffect}.

\section{Smearing kernel correction}
\label{app:smearingkernel}

The effect of the smearing on the expectation values of the Fourier mode amplitudes of interface fluctuations can be calculated semi-analytically. Let
\begin{equation}
A\of{x,y,z}=
\begin{cases}
A_0 \quad\text{if} & z_1\of{x,y} \leq z < z_2\of{x,y}\\
0 \quad\text{else} & \\
\end{cases}\ ,\label{eq:dcpsurfacerep}
\end{equation}
describe the shape of the volume occupied by the deconfined phase (cf. Fig.~\ref{fig:twophase}) with $z_i\of{x,y}$, $i=1,2$ being the parametrizations of the two interface surfaces. We neglect bulk fluctuations, so that $A_0=\abs{\avof{P_L}_{\text{deconf.}}}$ is constant. We now Fourier transform in $z$:
\begin{equation}
\hat{A}\of{x,y,k_z}=\frac{i\,A_0}{2\,\pi\,k_z}\,\of{e^{i\,k_z z_1\of{x,y}}-e^{i \,k_z z_2\of{x,y}}}\ .\label{eq:dcpsurfacerepftz}
\end{equation}
Since we neglect bulk fluctuations and the two phases occupy about the same space in z-direction in our setup, the small-$k_z$ modes should dominate in Eq.~\eqref{eq:dcpsurfacerepftz}. If we assume that the $z_i\of{x,y}$, $i=1,2$ are well separated, so that the amplitudes of their fluctuations as functions of $\of{x,y}$ are small relative to the wavelengths of the relevant $k_z$-modes, i.e.
\begin{equation}
z_i\of{x,y}=z_{0,i}+\Delta z_i\of{x,y}\quad \text{with}\quad \abs{\Delta z_i\,k_z}\ll 1\quad \forall\,i=1,2\quad ,\label{eq:dcpsurfaceparamappr}
\end{equation}
we can write
\begin{equation}
\hat{A}(x,y,k_z)=\frac{A_0}{2\,\pi}\,\of{-\frac{e^{i\,k_z z_{0,1}}-e^{i \,k_z z_{0,2}}}{i\,k_z} - e^{i\,k_z z_{0,1}}\,\Delta z_1\of{x,y} + e^{i\,k_z z_{0,2}}\,\Delta z_2\of{x,y}\,+\,\mathcal{O}\sof{k_z\,\Delta z_i^2\of{x,y}}}\ .\label{eq:dcpsurfacerepftzappr}
\end{equation}
Taking now the Fourier transform of $\tilde{A}(x,y,k_z)$ with respect to $x$ and $y$, using Eq.~\eqref{eq:dcpsurfacerepftzappr}, we find for $k_x^2+k_y^2>0$, i.e. the non-zero modes, that
\begin{equation}
\hat{A}\of{k_x,k_y,k_z}=\frac{A_0}{2\,\pi}\,\of{ - e^{i\,k_z z_{0,1}}\,\Delta \hat{z}_1\of{k_x,k_y} + e^{i\,k_z z_{0,2}}\,\Delta \hat{z}_2\of{k_x,k_y}\,+\,\mathcal{O}\of{k_z\,\Delta z_i^2}}\ ,\label{eq:dcpsurfacerepft}
\end{equation}
with $\Delta \hat{z}_i\of{k_x,k_y}$, $i=1,2$ being the Fourier transforms of the $\Delta z_i\of{x,y}$, $i=1,2$. Since the differences $z_i\of{x,y}-\Delta z_i\of{x,y}=z_{0,i}$, $i=1,2$ are constant, the Fourier transforms $\Delta \hat{z}_i\of{k_x,k_y}$, $i=1,2$ coincide for $k_x^2+k_y^2>0$ with the Fourier transforms $\hat{z}_i\of{k_x,k_y}$, $i=1,2$ of the interface surface parameterizations $z_i\of{x,y}$, $i=1,2$ and we therefore find that if Eq.~\eqref{eq:dcpsurfaceparamappr} is valid, we have
\begin{equation}
\hat{z}_i\of{k_x,k_y} \propto \hat{A}\of{k_x,k_y,k_z}\quad,\quad i=1,2\quad .\label{eq:dcpsurfacemodes}
\end{equation}

Now, because our "local order parameter" for deconfinement, i.e. the local Polyakov loop field $P_L\of{x,y,z}$ from Eq.~\eqref{eq:polyakovloop}, suffers from severe UV noise, Eq.~\eqref{eq:dcpsurfacerep} with $A_0$ constant is not a good approximation for $P_L\of{x,y,z}$ itself. However, with increasing number $n_s$ of smearing steps performed on $P_L\of{x,y,z}$ with the kernel Eq.~\eqref{eq:smearingkernel}, the resulting smoothed Polyakov loop field, $P^{\of{n_s}}_L\of{x,y,z}$ (cf. Fig.~\ref{fig:smearing_comp}), can be better and better represented by a correspondingly smoothed version of Eq.~\eqref{eq:dcpsurfacerep}. In Fourier space the latter is simply given by
\begin{equation}
\hat{A}^{(n_s)}\of{k_x,k_y,k_z} = \hat{S}^{n_s}\of{k_x,k_y,k_z}\,\hat{A}\of{k_x,k_y,k_z}\ ,\label{eq:dcpsurfacerepftsmeared}
\end{equation}
with
\begin{equation}
\hat{S}\of{k_x,k_y,k_z}=\frac{1+2\,\rho\,\of{\cos\of{k_x}+\cos\of{k_y}+\cos\of{k_z}}}{1+6\,\rho}=1-\rho\,\frac{4\of{\sin^2\of{k_x/2}+\sin^2\of{k_y/2}+\sin^2\of{k_z/2}}}{1+6\rho}\label{eq:smearingkernelft}
\end{equation}
being the Fourier transform of the smearing kernel from Eq.~\eqref{eq:smearingkernel}.
In the limit of small $k_z$ Eq.~\eqref{eq:dcpsurfacemodes} then suggests that
\begin{equation}
\hat{z}_i^{(n_s)}\of{k_x,k_y}=\hat{S}^{n_s}\of{k_x,k_y,k_z\approx 0}\,\hat{z}_i\of{k_x,k_y}\quad ,\quad i=1,2\quad .\label{eq:smearedunsmearedsurfacemoderel}
\end{equation}

Assuming that Eq.~\eqref{eq:dcpsurfacerepftsmeared} is indeed a good approximation for the Fourier transform of the smoothed Polyakov loop field $P^{\of{n_s}}_L\of{x,y,z}$, then Eq.~\eqref{eq:smearedunsmearedsurfacemoderel} suggests that
\begin{equation}
\savof{\abs{\hat{z}\of{k_x,k_y}}^2}=\frac{1}{\hat{S}^{2\,n_s}\of{k_x,k_y,k_z\approx 0}}\,\savof{\abs{\hat{z}^{\of{n_s}}\of{k_x,k_y}}^2}\ 
\end{equation}
can be used as a \emph{kernel-corrected} estimator for the interface surface fluctuation spectrum, which for sufficiently large $n_s$ becomes independent of the exact value of $n_s$.

Note that as long as $\rho<1$ the Fourier transformed smearing kernel $\hat{S}\of{k_x,k_y,k_z}$ can be written as:
\begin{equation}
\hat{S}\of{k_x,k_y,k_z}=\exp\sof{\log\sof{\hat{S}\of{k_x,k_y,k_z}}}=\exp\sbof{-\alpha_{\rho}\sum\limits_{\mathclap{\quad i\in\cof{x,y,z}}}\,\of{2\,\sin\of{k_i/2}}^2}+\order\sof{\alpha_{\rho}^2}\ ,
\end{equation}
with $\alpha_{\rho}=\rho/\of{1+6\,\rho}$ as in Eq.~\eqref{eq:plmodesmearingeffect}.

\section{Surface fluctuation spectrum without surface detection}
Consider again Eq.~\eqref{eq:dcpsurfacerep} from the previous section. Its partial Fourier transform in $z$, given in Eq.~\eqref{eq:dcpsurfacerepftz}, can also be written as
\begin{equation}
\hat{A}\of{x,y,k_z}=\frac{A_0}{\pi\,k_z}\,\sin\of{k_z\of{z_2\of{x,y}-z_1\of{x,y}}/2}\,\e^{i\,k_z\of{z_1\of{x,y}+z_2\of{x,y}}/2}\ .\label{eq:dcpsurfacerepftz2} 
\end{equation}
We can then define:
\begin{equation}
h\of{x,y}:=\arg\sof{\hat{A}\of{x,y,k_z}}/k_z=\of{z_1\of{x,y}+z_2\of{x,y}}/2\ ,
\end{equation}
where $k_z$ should be set to the lowest non-zero value. By taking the Fourier transform of $h\of{x,y}$, we find that its absolute value squared satisfies:
\begin{equation}
2\,\abs{\hat{h}\of{k_x,k_y}}^2=\frac{\abs{\hat{z}_1\of{k_x,k_y}}^2+\abs{\hat{z}_2\of{k_x,k_y}}^2}{2}+\mathrm{Re}\of{\hat{z}^{*}_1\of{k_x,k_y}\hat{z}_2\of{k_x,k_y}}\ ,\label{eq:hflucamplitude}
\end{equation}
where the last term should vanish in ensemble averages, provided that $z_1\of{x,y}$ and $z_2\of{x,y}$ fluctuate independently. 
One could therefore use
\begin{equation}
\avof{\abs{\hat{z}\of{k_x,k_y}}^2}=\frac{\avof{\abs{\hat{z}_1\of{k_x,k_y}}^2}+\avof{\abs{\hat{z}_2\of{k_x,k_y}}^2}}{2}=\savof{2\,\abs{\hat{h}\of{k_x,k_y}}^2}
\end{equation}
as estimator for the surface fluctuation spectrum. The advantage of this approach is that one does not have to actually locate the phase boundaries. However, smearing of the local Polyakov loop field would still be required to approach a situation where the system can be described by Eq.~\eqref{eq:dcpsurfacerep}.
\FloatBarrier

\section{Tables from interface tension analysis}\label{app:ittables}

\begin{table}[h]
    \small
    \centering
    \begin{tabular}{c | c | c | l | l | l | c }
        $N$  &  k.-corr.  &  $n_s$  &  \multicolumn{1}{c |}{$c_1$}  &  \multicolumn{1}{c |}{$c_2$}  &  \multicolumn{1}{c |}{$c_3$}  &  $\chi^2$/\#dof\\
        \hline\cmidrule{1-7}
        4  &  no  &  20   &  5.71(11)    & \phantom{-}0.308(16)   &  -0.0383(28)  &  3.792\\
           &      &  40   &  7.31(15)    & \phantom{-}0.132(17)   &  -0.0247(30)  &  1.597\\
           &      &  60   &  7.71(16)    & -0.014(17)   &  -0.0140(30)  &  1.047\\
           &      &  80   &  7.79(18)    & -0.146(18)   &  -0.0046(31)  &  0.737\\
           \cmidrule{2-7}
           & yes  &  20   &  5.71(11)    & \phantom{-}0.423(16)   &  -0.0384(28)  &  3.787\\
           &      &  40   &  7.30(15)    & \phantom{-}0.362(17)   &  -0.0247(30)  &  1.573\\
           &      &  60   &  7.71(16)    & \phantom{-}0.331(17)   &  -0.0141(30)  &  1.007\\
           &      &  80   &  7.78(17)    & \phantom{-}0.315(18)   &  -0.0048(31)  &  0.695\\
        \cmidrule{1-7}
        10 &  no  &  20   &  0.5397(75)  & \phantom{-}0.0607(84)  &  -0.0079(11)  &  1.070\\
           &      &  40   &  0.5404(76)  & -0.0682(85)  &  -0.0064(11)  &  1.462\\
           &      &  60   &  0.5397(77)  & -0.1913(86)  &  -0.0053(11)  &  1.809\\
           &      &  80   &  0.5397(77)  & -0.3110(86)  &  -0.0045(11)  &  2.155\\
           \cmidrule{2-7}
           & yes  &  20   &  0.5400(75)  & \phantom{-}0.1751(84)  &  -0.0078(11)  &  0.850\\
           &      &  40   &  0.5409(76)  & \phantom{-}0.1607(85)  &  -0.0062(11)  &  0.961\\
           &      &  60   &  0.5404(77)  & \phantom{-}0.1522(86)  &  -0.0052(11)  &  0.984\\
           &      &  80   &  0.5407(77)  & \phantom{-}0.1469(86)  &  -0.0043(11)  &  0.985\\
    \end{tabular}
    \caption{Results from the example zero-mode extrapolation fits shown in Fig.~\ref{fig:fouriermodes}, using the ansatz Eq.~\eqref{eq:spectrumfit} on kernel-corrected (k.-corr.: yes) and non-kernel-corrected (k.-corr.: no) interface fluctuation spectrum data for SU(4) at $\beta=10.94345$ on a $100^2 \times 400 \times 7$ lattice, and for SU(10) at $\beta=70.80712$ on a $80^2\times 240 \times 7$ lattice, obtained after $n_s=20,40,60,80$ smearing steps.}
    \label{tab:itzeromodeextrapfitres}
\end{table}

\begin{table}[h]
    \small
    \centering
    \begin{tabular}{r | c | c | c | c | c | c | c | l }
        $N$ & $N_t$ & $N_{x,y}$ & $N_z$ & $\beta$ & $n_s$ & $\big\lceil\frac{k^2}{T^2}\big\rceil$ & $\chi^2$/\#dof & \multicolumn{1}{ c }{$T_c^3/\sigma$}\\
    \hline \hline
   4 &    5 &   \phantom{0}80 &  320 &   \phantom{0}10.63790 &   \phantom{0}80   &     3.60     &    0.74    &       6.0(2)\\
     &    6 &   \phantom{0}80 &  200 &   \phantom{0}10.79192 &  100   &     4.28     &    0.11    &       7.2(2)\\
     &    7 &  100 &  400 &   \phantom{0}10.94345 &   \phantom{0}80   &     5.20     &    0.69    &       7.8(2)\\
     &    8 &  140 &  400 &   \phantom{0}11.08440 &  160   &     3.60     &    0.23    &       8.2(3)\\
\hline
   5 &    5 &   \phantom{0}80 &  240 &   \phantom{0}16.87620 &   \phantom{0}60   &     4.20     &    0.52    &      2.69(6)\\
     &    6 &  160 &  480 &   \phantom{0}17.11100 &   \phantom{0}50   &     3.92     &    0.78    &      3.13(3)\\
     &    7 &  100 &  400 &   \phantom{0}17.34282 &   \phantom{0}80   &     4.00     &    0.65    &      3.35(9)\\
     &    8 &  120 &  360 &   \phantom{0}17.56142 &  120   &     3.80     &    0.91    &      3.41(6)\\
\hline
   8 &    5 &   \phantom{0}50 &  180 &   \phantom{0}43.98370 &   \phantom{0}60   &     4.60     &    0.35    &      0.70(2)\\
     &    6 &   \phantom{0}80 &  240 &   \phantom{0}44.56196 &   \phantom{0}80   &     4.80     &    1.15    &      0.85(2)\\
     &    7 &   \phantom{0}80 &  260 &   \phantom{0}45.13525 &   \phantom{0}60   &     4.20     &    0.50    &      0.88(2)\\
     &    8 &  100 &  400 &   \phantom{0}45.67777 &   \phantom{0}60   &     4.80     &    1.07    &      0.91(2)\\
\hline
  10 &    5 &   \phantom{0}50 &  200 &   \phantom{0}69.03218 &   \phantom{0}40   &     5.40     &    0.32    &      0.44(2)\\
     &    6 &   \phantom{0}60 &  240 &   \phantom{0}69.92250 &   \phantom{0}70   &     5.60     &    0.51    &      0.51(2)\\
     &    7 &   \phantom{0}80 &  240 &   \phantom{0}70.80712 &   \phantom{0}80   &     5.80     &    0.98    &     0.541(8)\\
     &    8 &   \phantom{0}80 &  280 &   \phantom{0}71.64635 &  100   &     5.60     &    1.50    &     0.554(10)\\
\hline
  16 &    6 &   \phantom{0}40 &  160 &  179.85500 &   \phantom{0}40   &     7.60     &    0.55    &     0.179(5)\\
    \end{tabular}
    \caption{Table lists for each combination of $N$ and $N_t$ the values of $N_{x,y}$, $N_z$, and $\beta$ used to generate the surface fluctuation spectrum data for the interface tension analysis, together with the smearing level $n_s$ and maximum value of $k^2/T^2$ used to fit Eq.~\eqref{eq:spectrumfit} to the spectrum data, and the achieved $\chi^2/\text{\#dof}$ and fit result for $T_c^3/\sigma$.}
    \label{tab:sigma}
\end{table}

\begin{table}[h]
    \small
    \centering
    \begin{tabular}{l | c | c | c }
$N$ & $c_1$ & $c_2$ & $\chi^2$/\#dof\\
\hline\hline
4 &  0.00623(48) &  0.087(21) & 0.017\\
5 &  0.01033(44) &  0.087(17) & 0.321\\
8 &  0.01554(74) &  0.106(32) & 0.027\\
10 &  0.01619(84) &  0.117(40) & 0.129
    \end{tabular}
    \caption{Results from fitting Eq.~\eqref{eq:itcontinuumfit} to the interface tension data for $N=4\ldots 10$ and $N_t=6 \ldots 8$ from Table~\ref{tab:sigma} to obtain for each given $N$ the continuum value $\sigma/\of{T_c^3\,N^2}:=c_1$.}
    \label{tab:sigmacontinuumlim}
\end{table}

\begin{table}[h]
    \small
    \centering
    \begin{tabular}{l | c | l | l | c | c }
$a\,T_c$ & fit order & \multicolumn{1}{ c |}{$c_1$} & \multicolumn{1}{ c |}{$c_2$} & \multicolumn{1}{ c |}{$c_3$} & \multicolumn{1}{ c }{$\chi^2$/\#dof}\\
\hline\hline
$6^{-1}$ & linear &  0.02127(24) &  -0.2079(55) &  & 6.306\\
 & quadratic & 0.02246(42) & -0.279(22) & 0.93(27) & 0.965\\
\hline
$8^{-1}$ & linear &  0.02004(27) &  -0.2014(60) &  & 1.350\\
 & quadratic & 0.02058(47) & -0.242(30) & 0.54(39) & 0.785\\
\hline
0 & linear & 0.01824(67) &  -0.194(15) &  & 0.193\\
 & quadratic & 0.0186(12) & -0.220(71) & 0.35(92) & 0.244 \\
    \end{tabular}
    \caption{Large-$N$ extrapolation fit results, using the fit ans{\"a}tze from~Eq.~\eqref{eq:itlargeNfit} (linear) and Eq.~\eqref{eq:itlargeNfitquad} (quadratic) on the finite lattice spacing data for $\sigma/\of{T_c^3\,N^2}$ \refcorrb{as function of $1/N^2$} at $\of{a\,T_c}=1/N_t$, $N_t=6,8$ from Table~\ref{tab:sigma} and on the corresponding continuum extrapolated values form Table~\ref{tab:sigmacontinuumlim}. All fits have been carried out on data in the range $N\in\fof{4,10}$.}
    \label{tab:sigmalargeNlim}
\end{table}
\FloatBarrier

\section{Tables from latent heat analysis}\label{app:lhtables}
\begin{table}[h]
    \small
    \centering
    \begin{tabular}{l | l | l | l }
 $i$  &  $\ c_{i,0}$  &  $\ c_{i,1}$  &  $\ c_{i,2}$\\
\hline
 1  &   \ 3.48(4)  &  -6.8(5)  &   \ 7.3(15)\\
 2  &   \ 3.2(2)$\,10^1$  &  -1.6(3)$\,10^2$  &   \ 3.7(8)$\,10^2$\\
 3  &   \ 2.9(3)$\,10^2$  &  -4.9(32)$\,10^2$  &  -6.0(91)$\,10^2$\\
 4  &  -2.5(11)$\,10^3$  &  \ 6.4(14)$\,10^4$  &  -2.0(4)$\,10^5$\\
    \end{tabular}
    \caption{Obtained values for the parameters $c_{i,j}$ in the global ansatz \eqref{eq:interpg} for fitting the dependency of $a\,T_c$ on the inverse bare gauge coupling $\beta$ simultaneously for $N=4,5,8,10$ with a single function, using the procedure described below Eq.~\eqref{eq:interpg}. The final fit yield $\chi^2/\text{\#dof}=0.3$. Note that the magnitudes of the stated errors can be misleading due to considerable cross correlations among the 12 parameters.}
    \label{tab:betacglobalfitres}
\end{table}

\begin{table}[h]
    \small
    \centering
    \begin{tabular}{r | c | l | l | r | l | l | l | l }
 N  &  $N_t$  &  \multicolumn{1}{ c |}{$\beta_c$}  & \multicolumn{1}{ c |}{$-\frac{1}{N^2} \frac{\dd\beta}{\dd\log\of{a}}$}  &  \multicolumn{1}{ c |}{$N_s$}  &  \multicolumn{1}{ c |}{$\avof{U_{\Box}}_{\text{conf.}}$} & \multicolumn{1}{ c |}{$\avof{U_{\Box}}_{\text{deconf.}}$}  &  \multicolumn{1}{ c |}{$\Delta\avof{U_{\Box}}$}  &  \multicolumn{1}{ c }{$L/T_c^4$}\\
\hline
\hline
4  &  5  &   10.63777(9) &   0.0483(2)    &  80  &   0.4519797(10) &   0.450056(4)  &   0.001924(4)  &  \phantom{0}5.58(3)\\
   &  6  &   10.7919(2)  &   0.05736(6)   &  80  &   0.438034(2)   &   0.437355(2)  &   0.000679(3)  &  \phantom{0}4.84(2)\\
   &  7  &   10.9422(3)  &   0.06426(9)   &  100 &   0.4269259(5)  &   0.426619(2)  &   0.000307(3)  &  \phantom{0}4.55(4)\\
   &  8  &   11.0844(4)  &   0.0688(2)    &  80  &   0.4176187(7)  &   0.417455(2)  &   0.000164(3)  &  \phantom{0}4.44(7)\\
   & 10* &   11.3400(6)  &   0.0740(9)    &      &                 &                &                &  \\   
   & $\infty$ &          &                &      &                 &                &                &  \phantom{0}3.8(2)\\
\hline
5  &  5  &   16.8763(2)  &   0.04633(10)  &  80  &   0.461442(2)   &   0.459100(2)  &   0.002342(3)  &  10.17(3)\\
   &  6  &   17.1108(2)  &   0.05651(3)   &  80  &   0.446930(3)   &   0.4461147(9) &   0.000815(4)  &  \phantom{0}8.95(4)\\
   &  7  &   17.3427(3)  &   0.06333(3)   &  80  &   0.4354151(6)  &   0.4350449(7) &   0.000370(3)  &  \phantom{0}8.44(5)\\
   &  8  &   17.5612(2)  &   0.06738(6)   &  100 &   0.4258326(3)  &   0.4256337(5) &   0.000199(3)  &  \phantom{0}8.24(9)\\
   & 10* &   17.958(2)   &   0.0763(6)    &      &                 &                &                &  \\   
   & $\infty$ &          &                &      &                 &                &                &  \phantom{0}7.2(2)\\
\hline
8  &  5  &   43.9823(4)  &   0.04478(7)  &  60  &   0.471720(2)   &   0.469112(2)  &   0.002608(3)  &  28.03(5)\\
   &  6  &   44.5622(3)  &   0.05452(3)  &  80  &   0.4564330(5)  &   0.4555176(7) &   0.000915(3)  &  24.84(6)\\
   &  7  &   45.1355(3)  &   0.06138(3)  &  70  &   0.4444675(6)  &   0.4440437(7) &   0.000424(3)  &  24.0(2)\\
   &  8  &   45.6783(5)  &   0.06547(6)  &  100 &   0.4346367(4)  &   0.4344103(4) &   0.000226(3)  &  23.3(3)\\
   & 10* &   46.658(4)   &   0.0723(4)   &      &                 &                &                &  \\   
   & $\infty$ &          &               &      &                 &                &                &  21.5(4)\\
\hline
10 &  5  &   69.0313(4)  &   0.04438(8)  &  32  &   0.4740370(10) &   0.471380(3)  &   0.002657(4)  &  44.2(2)\\
   &  6  &   69.9245(4)  &   0.05366(3)  &  40  &   0.458589(2)   &   0.457648(2)  &   0.000941(3)  &  39.2(2)\\
   &  7  &   70.8076(10) &   0.06059(4)  &  60  &   0.4465565(4)  &   0.4461247(7) &   0.000432(3)  &  37.7(2)\\
   &  8  &   71.6476(10) &   0.06487(8)  &  40  &   0.4366563(3)  &   0.4364258(4) &   0.000230(3)  &  36.7(4)\\
   & 10* &   73.152(4)   &   0.0701(5)   &      &                 &                &                &  \\   
   & $\infty$ &          &               &      &                 &                &                &  33.5(6)\\
    \end{tabular}
    \caption{Table summarizing input data and results from the latent heat computation. The columns show, from left to right, the number of colors $N$, the temporal lattice size $N_t$, the critical value of the inverse gauge coupling $\beta_c$, the lattice beta function, the spatial lattice size used for the plaquette measurements, the average plaquette value at $\beta_c$ in the confined, and deconfined phase, as well as their difference, and the latent heat in units of $T_c^4$. The lattice beta function is shown in the lower panel of Fig.~\ref{fig:fitsforbetafunc} and was computed form the function Eq.~\eqref{eq:interpg}, fitted to the data $\cof{\of{\beta_c\of{N,N_t},N,-\log\of{N_t}}\vert N=4,5,8,10, N_t=5,6,7,8,10}$ from the first 3 columns with the method described below in Eq.~\eqref{eq:interpg}. The stated errors have been computed as in Eq.~\eqref{eq:errorbands}, i.e. by taking the full covariance matrix of the fit parameters into account. The obtained best fit parameter values are listed in Table~\ref{tab:betacglobalfitres}. Note, however, that the parameter errors stated in the latter table are misleading and insufficient to reproduce the errors in the lattice beta function, due to considerable cross correlations among the fit parameters. The errors of the plaquette differences include an additional error contribution of order $2\times 10^{-6}$, taking into account the effect of non-zero topological charges on the average plaquette values in the confined phase (cf. Fig.~\ref{fig:plaqvstopo}). The asterisk for $N_t=10$ indicates that these values for $\beta_c$ have been obtained on significantly smaller spatial lattices.}
    \label{tab:latentheatcompdat}
\end{table}

\begin{table}[h]
    \small
    \centering
    \begin{tabular}{r | c | c | c }
N & \multicolumn{1}{ c |}{$c_1$} & \multicolumn{1}{ c |}{$c_2$} & \multicolumn{1}{ c }{$\chi^2$/\#dof}\\
\hline\hline
4 &  0.2392(68) &  2.28(26) & 0.858\\
5 &  0.2875(63) &  2.53(26) & 1.001\\
8 &  0.3356(56) &  1.89(21) & 0.181\\
10 &  0.3347(57) &  2.08(23) & 0.013\\
    \end{tabular}
    \caption{Results from fitting Eq.~\eqref{eq:itcontinuumfit} to the latent heat data for $N=4\ldots 10$ and $N_t=6 \ldots 8$ from Table~\ref{tab:latentheatcompdat}, to obtain for each given $N$ the continuum value $L/\of{N^2\,T_c^4}:=c_1$.}
    \label{tab:latheatcontinuumlim}
\end{table}

\begin{table}[h]
    \small
    \centering
    \begin{tabular}{l | c | l | l | l | c }
$a\,T_c$ & fit order & \multicolumn{1}{ c |}{$c_1$} & \multicolumn{1}{ c |}{$c_1$} & \multicolumn{1}{ c |}{$c_1$} & \multicolumn{1}{ c }{$\chi^2$/\text{\#dof}}\\
\hline\hline
$6^{-1}$ & linear &  0.414(7) &  -1.71(19) &  & 44.0\phantom{0}\\
 & quadratic & 0.393(3) & \phantom{-}0.08(19) & -24.5(27) & \phantom{0}0.87\\
\hline
$8^{-1}$ & linear &  0.389(7) &  -1.67(20) &  & \phantom{0}4.8\phantom{0}\\
 & quadratic & 0.373(6) & -0.27(46) & -20.0(64) & \phantom{0}0.07\\
\hline
0 & linear & 0.360(6) &  -1.88(17) &  & \phantom{0}1.26\\
 & quadratic & 0.349(12) & -0.97(92) & -12.8(128) & \phantom{0}1.25 \\
    \end{tabular}
    \caption{Large-$N$ extrapolation fit results, using the fit ans{\"a}tze from~Eq.~\eqref{eq:itlargeNfit} (linear) and Eq.~\eqref{eq:itlargeNfitquad} (quadratic) on the data for $L/\of{N^2\,T_c^4}$ \refcorrb{as function of $1/N^2$} at $\of{a\,T_c}=1/N_t$, $N_t=6,8,\infty$ from Table~\ref{tab:latentheatcompdat}.}
    \label{tab:lhlargeNlim}
\end{table}

\FloatBarrier
\end{widetext}

\bibliography{interface_tension_and_latent_heat_at_large_Nc}

\end{document}